\documentclass{jfm}
\usepackage{graphicx}
\usepackage{epstopdf, epsfig}

\usepackage{mathtools,amsmath}
\usepackage{xcolor}
\usepackage{makecell,booktabs,colortbl,array}
\usepackage{pifont}
\usepackage[colorlinks=true,citecolor=blue]{hyperref}

\shorttitle{Liouville chains of hybrid vortex equilibria}
\shortauthor{V. S. Krishnamurthy, M. H. Wheeler, D. G. Crowdy and A. Constantin}

\title{Liouville chains: new hybrid vortex equilibria of the 2D Euler equation}

\author{Vikas S. Krishnamurthy\aff{1}
  \corresp{\email{vikas.krishnamurthy2@gmail.com}},
  Miles H. Wheeler\aff{2},
  Darren G. Crowdy\aff{3} 
  \and 
  Adrian Constantin\aff{1}}

\affiliation{\aff{1}Faculty of Mathematics, University of Vienna, Oskar-Morgenstern-Platz 1, 1090 Vienna, Austria
	\aff{2}Department of Mathematical Sciences, University of Bath, Bath, BA2 7AY, UK
\aff{3}Department of Mathematics, Imperial College London, 180 Queen's Gate, London SW7 2AZ, UK}

\begin{document}

\maketitle

\begin{abstract}
	A large class of new exact solutions to the steady, incompressible Euler equation on the plane is presented. 
	These hybrid solutions consist of a set of stationary point vortices embedded in a background sea of Liouville-type vorticity that is exponentially related to the stream function.
	The input to the construction is a ``pure'' point vortex equilibrium in a background irrotational flow.  
	Pure point vortex equilibria also appear as a parameter $A$ in the hybrid solutions approaches the limits $A\to 0,\infty$.
	While $A\to0$ reproduces the input equilibrium, $A\to\infty$ produces a new pure point vortex equilibrium.
	We refer to the family of hybrid equilibria continuously parametrised by $A$ as a ``Liouville link''.
	In some cases, the emergent point vortex equilibrium as $A\to\infty$ can itself be the input for a second family of hybrid equilibria linking, in a limit, to yet another pure point vortex equilibrium.
	In this way, Liouville links together form a ``Liouville chain''.
	We discuss several examples of Liouville chains and demonstrate that they can have a finite or an infinite number of links.
	We show here that the class of hybrid solutions found by \citet{Crowdy2003a} and by \citet{KWCC2019} form the first two links in one such infinite chain.
	We also show that the stationary point vortex equilibria recently studied by \cite{KWCC2020} can be interpreted as the limits of a Liouville link.
	Our results point to a rich theoretical structure underlying this class of equilibria of the 2D Euler equation.

\end{abstract}


\section{Introduction}
A selection of exact solutions to the Navier--Stokes equation is provided by \cite{Drazin2006}.
Discussing the meaning of exact solutions they observe that ``it often denotes a solution which has a simple explicit form, usually an expression in finite terms of elementary or other well known special functions''.
This statement also applies to the Euler equation, the inviscid version of the Navier--Stokes equation.
In this paper, we present highly non-trivial exact vortex solutions of the steady Euler equation, but simply given in terms of rational functions.

The foundations of vortex dynamics were laid by~\cite{Helmholtz1858}, and since then various exact vortex solutions of the Euler equation have been described.
The importance of cataloging such solutions, and their applicability, is discussed in \citet{Saffman1981a}.
Many of the basic solutions are found in the classic textbooks of fluid dynamics~\citep{Lamb1993,Saffman1992}.
Examples of inviscid vortices in two-dimensions include point vortices and finite-area vortices such as the Rankine vortex, the Kirchhoff ellipse and the Lamb vortices. 
Kirchoff vortices are connected to Gerstner's famous wave solutions and other explicit solutions in Lagrangian coordinates \citep{AY1985}.

Point vortex configurations in which the vortices are stationary with respect to each other are called relative equilibria or vortex crystals.
\emph{Stationary equilibria} are a subset in which all the vortices are at complete rest.
The most basic point vortex equilibrium is an isolated point vortex which remains stationary. 
A vortex pair either translates or rotates uniformly depending on whether or not the total vortex circulation is zero. 
For $M\geq 3$ vortices, only a subset of possible motions are relative equilibria~\citep{Newton2001}, many of which remain to be described, even for small $M$ \citep{ONeil1987a}.
See~\cite{Aref2003} for a review of relative equilibria.

The notion of ``hybrid vortex equilibria'', in which two or more basic vortex models are taken to exist together in equilibrium, has been the topic of extensive study in recent years. 
\citet{CrowdyMP1} found a class of analytical solutions in which a vortex patch exists in equilibrium with a finite distribution of point vortices. 
His construction is based on posing the stream function in the form of a so-called modified Schwarz potential, an idea that has proven to be fruitful in generating large classes of equilibria, including solutions involving multiple patches \citep{CrowdyMP2, CrowdyMP3}. 
In a similar vein \citet{ONeil2018,ONeil2018b} has found solutions whereby a vortex sheet sits in equilibrium with an array of point vortices, a construction that also appears to yield many new equilibrium solutions.

The vortex solutions discussed above have compact regions of vorticity surrounded by irrotational flow. 
They are of particular interest due to the well-known presence of coherent structures in a variety of high Reynolds number flows.
In the context of his studies on mixing layers, \cite{Stuart1967} considered steady solutions to the planar Euler equation in which the flow is everywhere rotational.
The vorticity $\zeta$ is taken to be the exponential of the stream function $\psi$, i.e. $\zeta=-a\exp(b\psi)$, where $a$ and $b$ are real constants.
The resulting differential equation 
\begin{gather}\label{eq:liouville}
	\nabla^2\psi = a\exp(b\psi)
\end{gather}
is the semi-linear Liouville partial differential equation for the stream function.
The general solution of the Liouville equation can be written down explicitly in terms of a complex-analytic function $h(z)$ with isolated simple pole singularities \citep{Crowdy1997} and is given by
\begin{gather}\label{eq:liouville-sol}
	\psi(z,\bar{z}) = \frac{1}{b}\log\left[\frac{2|h'(z)|^2}{-ab(1+|h(z)|^2)^2}\right],
	\quad ab<0,
\end{gather}
where the two-dimensional fluid flow is taken to be in the complex $z$-plane, primes denote derivatives with respect to the argument, overbar denotes a complex conjugate and $|\cdot|$ denotes the modulus of a complex number.
\cite{Stuart1967} explored some solutions of the Liouville equation \eqref{eq:liouville} and identified a particular class which has become known in the fluid dynamics community as  Stuart vortices \citep{Saffman1992}.

Stuart vortices are closely related to solutions with compact vorticity surrounded by irrotational flow; indeed \cite{Stuart1967} shows that his solutions can be continuously varied from a tanh velocity profile, through Stuart vortices, to a limiting case of a point vortex row in otherwise irrotational flow.
It must be noted that, in the non-limiting case, Stuart vortices are everywhere smooth solutions of the Euler equation and exhibit the famous Kelvin's cat's eye streamline patterns.

A natural question then arises: can a set of point vortices be superimposed on a smooth background Stuart-type vorticity field to produce hybrid equilibria which are steady solutions of the Euler equation?
\cite{Crowdy2003a} first proposed an extension of Stuart's model to the case of hybrid vortex equilibria in which a steady point vortex exists in a smooth ambient background field of non-zero vorticity.
For an integer $N \ge 2$ he found a class of $N$-fold symmetric solutions for a central point vortex surrounded by a continuous non-zero distribution of Stuart-type vorticity---or what we will call henceforth, in view of the connection to \eqref{eq:liouville},  ``Liouville-type'' vorticity---having $N$ vortices with smoothly distributed (i.e. non-singular) vorticity.
Considering a certain limit of the solutions, he showed that they reduce to an axisymmetric flow with a single point vortex at the origin.
He further showed that, in another limit, the solutions become the $N$-fold symmetric pure point vortex equilibria studied earlier by \cite{Morikawa}, comprising $N$ point vortices in a polygonal arrangement with a central point vortex still present at the origin.

Since the work by \citet{Crowdy2003a}, hybrid solutions containing point vortices embedded in a Liouville-type background have been developed in various directions.
The planar solutions of \citet{Crowdy2003a} were later rediscovered by \cite{Tur2004}, also see \cite{Tur2011}.
Generalising the planar Stuart vortices to the case of a non-rotating sphere, \cite{Crowdy2004b} found analytical solutions for everywhere smooth vorticity on the surface of a sphere except for point vortices at the north and south poles.
These ideas can be extended to obtain Stuart vortex solutions on a torus \citep{Sakajo2019} and on a hyperbolic sphere \citep{Yoon2020}.
The introduction in \citet{KWCC2019} discusses other applications and extensions of Stuart vortices.

The above mentioned studies of hybrid equilibria contain either a single point vortex in the plane or two point vortices on compact surfaces.
In these cases, invoking symmetry arguments is sufficient to ensure that the point vortices are stationary, and the solutions obtained are therefore steady.
In a recent paper \cite{KWCC2019} showed the existence of an asymmetric family of hybrid vortex equilibria (although the background field is referred to there as ``Stuart-type'' vorticity in deference to \cite{Stuart1967}). 
They showed that a colinear three point vortex equilibrium, which is a limiting case    of the $N=2$ hybrid equilibrium discussed in \cite{Crowdy2003a}, can be continuously deformed into another non-trivial family of hybrid equilibria comprising a point vortex pair in equilibrium in an ambient field of Liouville-type vorticity. 
A certain limit of these hybrid equilibria produce another pure point vortex equilibrium in which the point vortex pair sit in equilibrium with eight other point vortices of opposite-signed circulation. 

We now briefly mention some relevant connections between pure point vortex equilibria and areas of mathematical physics.
The interested reader may consult \cite{Aref2007b,Aref2007,Aref2011a} and \cite{Clarkson2009} for a general discussion.
\citet{Burchnall1930} related the question of finding rational anti-derivatives of rational functions to the existence of polynomial solutions to a certain ``bilinear differential equation'' and constructed a Wronskian representation for the polynomials. 
These polynomials, called Adler--Moser polynomials, also arose in the context of describing rational solutions to the Korteweg--de Vries equation~\citep{Airault1977}.
They were constructed using iterated Darboux--Crum transformations of Schr\"odinger operators by \citet{Adler1978}.

The bilinear differential equation is called Tkachenko's equation in the vortex dynamics literature \citep{Aref2003,Tkachenko1964}.
\cite{Bartman1984} showed that point vortices of the same circulation but mixed sign, located at the roots of the Adler--Moser polynomials, are in stationary equilibrium.
Point vortices with circulation ratios $-2$ and $-1/2$ can also be in stationary equilibrium, if they are located at the roots of polynomials in two new hierarchies found by \citet{Loutsenko2004}, who studied a generalisation of the underlying bilinear differential equation.
For a more detailed discussion of these matters, the reader is referred to \citet{KWCC2020}.
A new transformation is introduced there that takes a given stationary point vortex equilibrium and produces a new stationary point vortex equilibrium. 
It presents a unified approach to obtain both the Adler--Moser polynomials and the Loutsenko polynomials using this new transformation.
Also see \citet{ONeil2006}, who studies continuous families of point vortex equilibria.

In the present paper, we consider solutions of the Liouville-type equation
\begin{gather}\label{eq:liouville-type}
	\nabla^2\psi = a\exp(b\psi) - \sum_{j=1}^{\widetilde{M}}\widetilde\Gamma_j\,\delta(z-\widetilde{z}_j),
\end{gather}
where in addition to the Liouville-type background vorticity there are $\widetilde{M}$ point vortices located at $\widetilde{z}_j$ and with circulations $\widetilde\varGamma_j$.
The point vortices are stationary if and only if the local expansions of the 2D fluid velocity field $(u,v)$ are of the form
\begin{gather}\label{eq:pv-stationary}
u-\mathrm{i}v = 2\mathrm{i}\frac{\partial\psi}{\partial z} = \frac{1}{2\upi\mathrm{i}}\frac{\widetilde\Gamma_k}{z-\widetilde{z}_k} + O(|z-\widetilde{z}_k|),
\end{gather}
for $k=1,2,\ldots,\widetilde{M}$.
Away from the point vortices, \eqref{eq:liouville-type} is still solved by \eqref{eq:liouville-sol}, but it is necessary that the point vortices are stationary in order to obtain steady solutions of the 2D Euler equation.
When multiple point vortices are present in the flow, non-trivial arguments are needed to show that they are all stationary.

\begin{figure}
	\centering
	\includegraphics[scale=0.8]{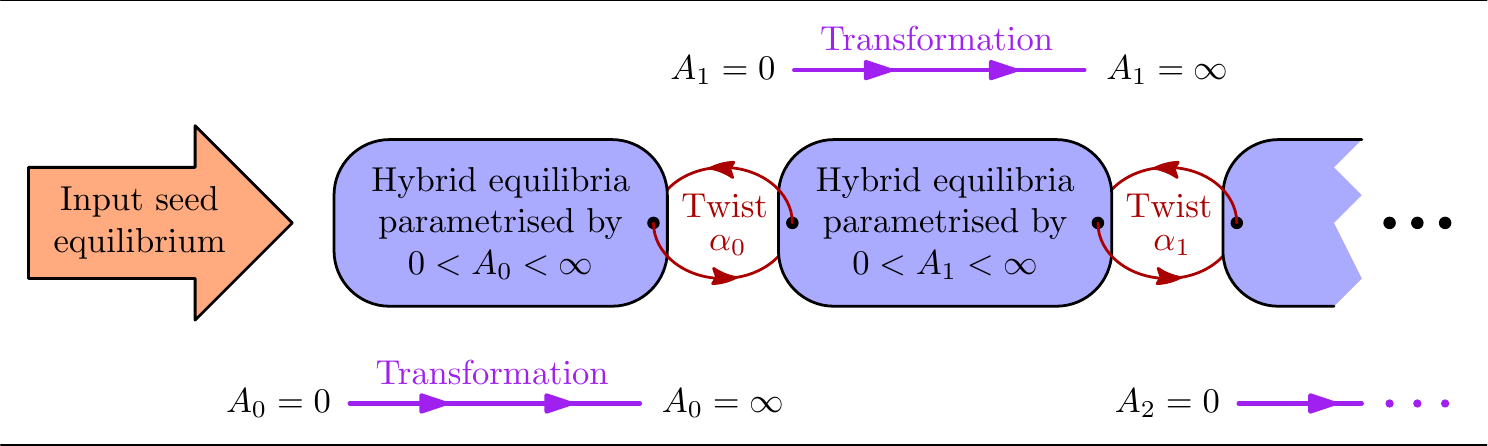}
	\caption{ Schematic of a ``Liouville chain''.
		A chain begins with a simple ``seed'' equilibrium, such as a single isolated point vortex in otherwise irrotational flow.
		Each link of the chain, called a ``Liouville link'', is a continuum of hybrid equilibria parameterized by some $A_n> 0$ ($n\geq 0$).
		The points where links connect, corresponding to $A_n = \infty$ and $A_{n+1} = 0$, are pure point vortex equilibria (i.e. in an		irrotational background). 
		There also exists a transformation, introduced in \cite{KWCC2020}, that allows jumping directly between the pure point vortex equilibria at the end points of a Liouville link, without having to ``pass through'' the intermediate hybrid equilibria.
		A ``twist'', quantified by a ``twist parameter'' $\alpha_n$ in our construction, is needed at each pure point vortex equilibrium to build the next link in the chain, or to jump to the next end point in the chain. 
		We present examples of single-link (\S\ref{sec:single}), $N$-link (\S\ref{sec:N}) and infinite (\S\ref{sec:infinite}) Liouville chains in this paper.
		A detailed worked example is presented in \S\ref{sec:example}. }
	\label{fig:schematic}
\end{figure}

The hybrid solutions of \citet{Crowdy2003a} correspond to $\widetilde{M}=1$ in \eqref{eq:liouville-type}, and invoking symmetry is sufficient to satisfy \eqref{eq:pv-stationary}.
The family studied by \citet{KWCC2019} corresponds to the case $\widetilde{M}=2$.
These families of hybrid equilibria and their point vortex limits signal the possibility that certain point vortex equilibria can be ``connected'' by families of hybrid equilibria. 
A key contribution of the present paper is to show the existence of a remarkably broad class of hybrid equilibria, including chains of hybrid solutions extrapolating between pure point vortex equilibria. 
We refer to these as ``Liouville chains'' since the extrapolating hybrid solutions---which we think of metaphorically as connected ``Liouville links'' in a chain---involve an exponential vorticity-stream function relation of Liouville-type {\it viz.} \eqref{eq:liouville-type}.
Figure~\ref{fig:schematic} shows a schematic of a Liouville chain, which can be finite or infinite in length. 
A mathematical ``twist'' is needed at a point vortex equilibrium connecting the links in the chain to continue on to the next link; this is encoded in a ``twist parameter'' $\alpha$. 
In our chain metaphor we call this mathematical operation a ``twist'' in analogy with the fact, as shown schematically in figure~\ref{fig:schematic}, that neighbouring links in a chain have to be rotated (``twisted'') in order to properly fit together.

We study three infinite Liouville chains in this paper. 
One of the infinite Liouville chains unveiled here has mathematical connections with the Adler--Moser polynomials \citep{Adler1978} discussed earlier. 
The two different polynomial hierarchies described by \citet{Loutsenko2004} in connection with his studies on the equilibria of Coulomb gases are also shown to have associated infinite Liouville chains of hybrid equilibria. 
While it is well-known that the Adler--Moser polynomials have a connection to point vortex dynamics \citep{Aref2003,Clarkson2009}, this paper shows for the first time that they can be used to represent much more complex hybrid vortical equilibria involving distributed vorticity of Liouville-type.
The families of hybrid equilibria studied by \citet{Crowdy2003a} and \citet{KWCC2019} are just two Liouville links in one of the examples studied here, namely, the infinite Liouville chain given in terms of the first hierarchy of Loutsenko polynomials.

This paper is organised as follows. 
In \S\ref{sec:pv-sv} we recall in detail point vortices, Stuart vortices \citep{Stuart1967} and the polygonal solutions of \citet{Crowdy2003a}.
Mathematical statements of the main results obtained in this paper are given in \S\ref{sec:results}.
A detailed example describing the construction procedure for a Liouville chain is given in \S\ref{sec:example}.
The general theory and justification for the results stated in \S\ref{sec:results} are provided in \S\ref{sec:theory-links} and \S\ref{sec:theory-chains}.
Many examples of the theory are presented in the subsequent sections: examples of single-link Liouville chains are given in \S\ref{sec:single}, $N$-link Liouville chains in \S\ref{sec:N} and infinite Liouville chains in \S\ref{sec:infinite}.
We summarise and discuss possible future directions in \S\ref{sec:summary}.

\section{Background theory and examples}\label{sec:pv-sv}
We consider the two-dimensional flow of an incompressible, inviscid and homogeneous fluid.
The incompressibility condition allows us to introduce the stream function $\psi$, up to an additive constant, by
\begin{gather}\label{eq:def-sf}
u = \frac{\partial\psi}{\partial y}
\quad\text{and}\quad 
v = -\frac{\partial\psi}{\partial x}.
\end{gather}
Here $(x,y)$ are the Cartesian coordinates of a planar cross-section of the flow, and $(u,v)$ are the components of the fluid velocity.
The vorticity has a single non-zero component 
\begin{gather}\label{eq:vort-sf}
\zeta = \frac{\partial v}{\partial x} - \frac{\partial u}{\partial y} = -\nabla^2\psi,
\end{gather}
where $\nabla^2=\partial^2/\partial x^2 + \partial^2/\partial y^2$ is the planar Laplacian operator.

The vorticity equation written in terms of the stream function is then~\citep{Saffman1992,Newton2001} 
\begin{gather}\label{eq:vorteq-sf}
\frac{\partial(\nabla^2\psi)}{\partial t} + \frac{\partial\psi}{\partial y}\frac{\partial(\nabla^2\psi)}{\partial x}
- \frac{\partial\psi}{\partial x}\frac{\partial(\nabla^2\psi)}{\partial y} = 0.
\end{gather}
It can be checked that any smooth stream function $\psi$ satisfying an equation of the form
\begin{gather}\label{eq:steady-vorteq}
\nabla^2\psi = V(\psi)
\end{gather}
is a steady solution of \eqref{eq:vorteq-sf} with vorticity $\zeta=-V(\psi)$.
Here $V(\psi)$ is any differentiable function.

It is convenient to work in the complex flow plane $z=x+\mathrm{i}y$, which can be related to the Cartesian plane via the formal change of variables $(x,y)\mapsto(z,\bar{z})$.
Here $\bar{z}=x-\mathrm{i}y$ and overbars denote complex conjugation.

\subsection{Point vortices}\label{ss:pv}
Point vortices have a vorticity distribution of the form~\citep{Newton2001}
\begin{gather}\label{eq:pv-vort}
\zeta = -\nabla^2\psi = \sum_{j=1}^M\varGamma_j\,\delta(z-z_j),
\end{gather}
where $z_j=x_j+\mathrm{i}y_j$ are the time-dependent locations of the point vortices and the constants $\varGamma_j$ are the circulations or strengths of the point vortices.
The corresponding stream function is:
\begin{gather}\label{eq:pv-sf}
\psi(z,\bar z) = -\frac{1}{2\upi}\sum_{j=1}^M\varGamma_j\log|z-z_j|,
\end{gather}
which is the imaginary part of the complex potential 
\begin{gather}\label{eq:pv-complexpot}
f(z) = \frac{1}{2\upi\mathrm{i}}\sum_{j=1}^M\varGamma_j\log(z-z_j).
\end{gather}
The velocity field due to the point vortices is given in terms $f(z)$ as
\begin{gather}\label{eq:pv-vel}
u-\mathrm{i}v
= f'(z) 
= \frac{1}{2\upi\mathrm{i}}\sum_{j=1}^M\frac{\varGamma_j}{z-z_j}.
\end{gather}
The complex potential \eqref{eq:pv-complexpot} and the complex velocity \eqref{eq:pv-vel} are complex-analytic functions of $z$ alone, independent of $\bar{z}$.
The velocity of a point vortex is obtained from \eqref{eq:pv-vel} after subtracting off the singular term and evaluating at the point vortex location:
\begin{gather}\label{eq:pv-nsi-vel}
\overline{\frac{\mathrm{d}z_k}{\mathrm{d}t}} 
= \left[(u-\mathrm{i}v)-\frac{\varGamma_k}{2\upi\mathrm{i}}\frac{1}{z-z_k}\right]\Bigg|_{z=z_k}
= \frac{1}{2\upi\mathrm{i}}\sum_{\genfrac{}{}{0pt}{}{j=1}{j\neq k}}^M\frac{\varGamma_j}{z_k-z_j},
\end{gather}
for $k=1,2,\ldots,M$~\citep{Saffman1992,Newton2001,LlewellynSmith2011}. 

In this paper we study stationary configurations of point vortices in which all vortex velocities are zero.
Then \eqref{eq:pv-nsi-vel} reduces to the $M$ algebraic conditions on the $M$ unknown vortex positions $z_1,\ldots,z_M$:
\begin{gather}\label{eq:pv-stat}
\sum_{\substack{j=1\\ j\neq k}}^M\frac{\varGamma_j}{z_k-z_j}=0\quad\text{for } k=1,2,\ldots,M,
\end{gather}
for given values of the circulations $\varGamma_1,\ldots,\varGamma_M$.
For further discussion of point vortex equilibria see \cite{KWCC2020} and the review article \cite{Aref2003}.

\subsection{Stuart vortices }\label{ss:Sv}
\begin{figure}
	\centering
	\includegraphics[scale=0.5]{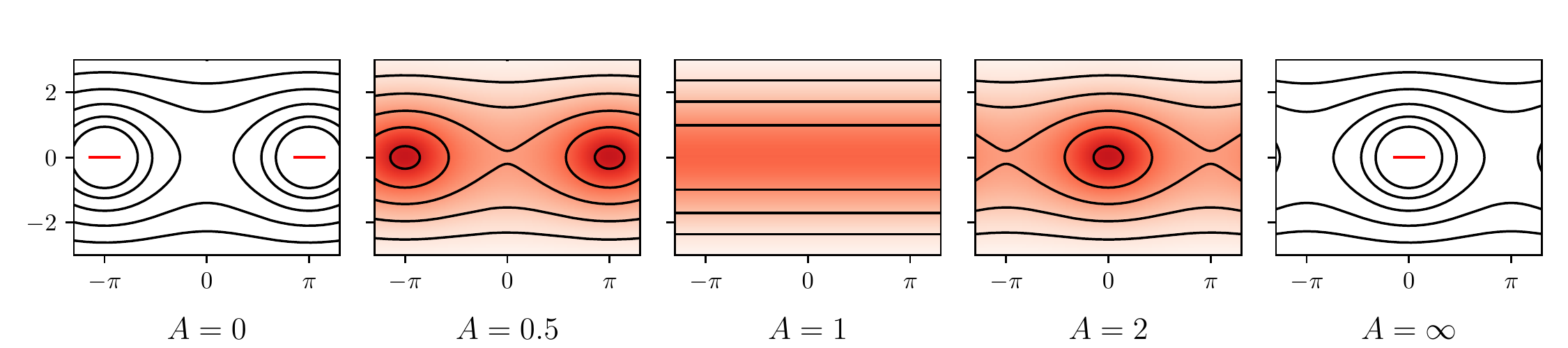}
	\caption{Streamlines and vorticity for Stuart vortices \citep{Stuart1967} given by the stream function \eqref{eq:stuart-sf}.
		Panels with a white background show streamline patterns for point vortices with negative ({\color{red}$-$}) circulation in otherwise irrotational flow. 
		The middle panels show the everywhere rotational and smooth flow for finite $A\neq 0$.
		In the limiting cases $A=0,\infty$, the smooth vorticity concentrates into a periodic row of point vortices with complex potentials \eqref{eq:stuart-cp}, surrounded by irrotational flow.}
	\label{fig:stuart}
\end{figure}
\cite{Stuart1967} considered steady solutions by choosing $V(\psi) = \exp(-2\psi)$ in \eqref{eq:steady-vorteq}.
In this case we get the Liouville equation \eqref{eq:liouville} with $a=1$ and $b=-2$, the general solution of which is given by \eqref{eq:liouville-sol}.
Then, in our present notation, Stuart's original solution is obtained by substituting
\begin{gather}\label{eq:stuart-sol}
	h(z) = A\tan\left(\frac{z}{2}\right) 
	\quad\implies\quad
	h'(z) = \frac{A}{2}\sec^2\left(\frac{z}{2}\right)
\end{gather}
in \eqref{eq:liouville-sol}.
Here $A$ is a real constant.
Subtracting an unimportant constant leads to the stream function
\begin{gather}\label{eq:stuart-sf}
\psi(z,\bar{z};A) = \log\left[\frac{1}{A}\left|\cos^2\left(\frac{z}{2}\right)\right|
+ A\left|\sin^2\left(\frac{z}{2}\right)\right|\right],
\end{gather}
which is smooth for all finite values of $A$ due to the fact that $h'(z)$ does not vanish anywhere in the complex plane.

\begin{subequations}
By taking appropriate limits as $A\to 0,\infty$ in \eqref{eq:stuart-sf}, we get the limiting stream functions 
\begin{align}
	\lim_{A\to 0}\left[\psi+\log A\right] &= 2\log\left|\cos\left(\frac{z}{2}\right)\right|,\\
	\lim_{A\to\infty}\left[\psi-\log A\right] &= 2\log\left|\sin\left(\frac{z}{2}\right)\right|.
\end{align}
\end{subequations}
These limiting stream functions are the imaginary parts of the complex potentials
\begin{gather}\label{eq:stuart-cp}
	G(z) = 2\mathrm{i}\log\cos\left(\frac{z}{2}\right)
	\quad\text{and}\quad
	F(z) = 2\mathrm{i}\log\sin\left(\frac{z}{2}\right).
\end{gather}
Both the complex potentials $G(z)$ and $F(z)$ correspond to an infinite row of point vortices with circulations $-4\upi$ each and consecutive vortices separated by a distance $2\upi$~\citep{Saffman1992}.
Figure~\ref{fig:stuart} shows streamline plots for the Stuart vortices and their limiting cases.

\subsection{Polygonal $N$-vortex equilibria}
\begin{figure}
	\centering
	\includegraphics[scale=0.5]{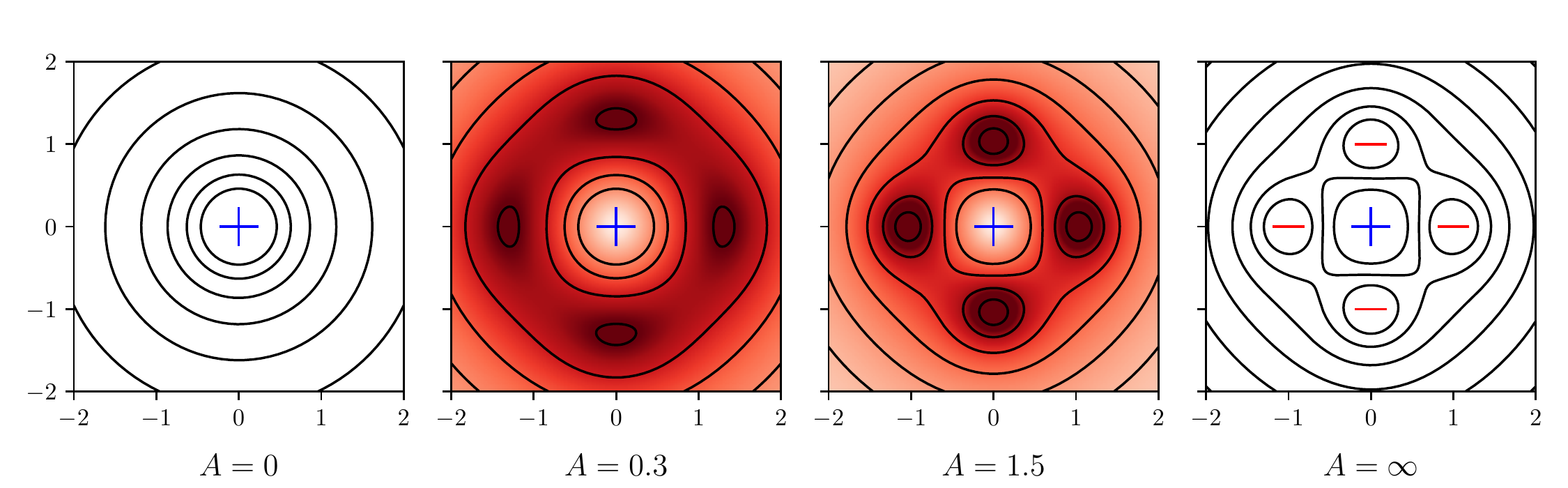}
	\caption{Panels with a white background show streamline patterns for point vortices with positive ({\color{blue}$+$}) and negative ({\color{red}$-$}) circulation in otherwise irrotational flow. 
	The middle panels show point vortices embedded in a sea of smooth Liouville-type background vorticity, which is negative and shaded in red.
	Streamlines and vorticity are shown for the $N$-polygonal equilibria \citep{Crowdy2003a} given by the stream function \eqref{eq:sf-poly}.
	The limit $A\to 0$ is an isolated point vortex, whereas the limit $A\to\infty$ is a centered polygon of stationary point vortices in other irrotational flow.}
	\label{fig:polygon}
\end{figure}
\citet{Stuart1967} made the choice \eqref{eq:stuart-sol} since he was interested in obtaining everywhere smooth solutions, i.e. without point vortex singularities.
\citet{Crowdy2003a} instead considered the function (rewritten in our notation)
\begin{gather}\label{eq:g-poly}
	h(z) = A(z^N+C)
	\quad\implies\quad
	h'(z) = ANz^{N-1},
\end{gather}
for integers $N\geq 2$, and where $A$ is a real constant and $C$ is a complex constant.
Substituting \eqref{eq:g-poly} into \eqref{eq:liouville-sol}, again with $a=1$ and $b=-2$, we get the stream function
\begin{gather}\label{eq:sf-poly}
	\psi(z,\bar{z};A,C) = \log\left[\frac{1}{A}\frac{1}{|z|^{N-1}}+A\frac{|z^N+C|^2}{|z|^{N-1}}\right],
\end{gather}
where we have dropped an unimportant constant $\log N$.
The hybrid stream function \eqref{eq:sf-poly} shows a point vortex singularity at $z_1=0$ with strength $\varGamma_1=2\upi(N-1)/2$.
The point vortex at the origin is surrounded by $N$ smooth vortices arranged on a regular polygon as shown in figure~\ref{fig:polygon} for the case $N=4$.
The symmetry of the solution guarantees that the point vortex at the origin in the hybrid solution \eqref{eq:sf-poly} remains stationary according to \eqref{eq:pv-stationary}.

Just as was done for \eqref{eq:stuart-sf}, it is useful to look at the two limits ${A\to 0}$ and ${A\to\infty}$ of \eqref{eq:sf-poly}.
We find
\begin{subequations}\label{eq:sf-limit-poly}
\begin{align}
	\lim_{A\to 0}\left[\psi+\log A\right] 
	&= -\log\left|z\right|^{N-1},\\
	\lim_{A\to\infty}\left[\psi-\log A\right] 
	&= -\log\frac{|z|^{N-1}}{|z^N+C|^2}.
\end{align}
\end{subequations}
We emphasize that while the stream function \eqref{eq:sf-poly} is \emph{not} harmonic, both the limits in \eqref{eq:sf-limit-poly} are harmonic.
Indeed, the limits are imaginary parts of the respective complex potentials
\begin{gather}\label{eq:limit-poly}
	G(z) = \frac{\varGamma_1}{2\upi\mathrm{i}}\log z
	\quad\text{and}\quad
	F(z;C) = \frac{\varGamma_1}{2\upi\mathrm{i}}\log{z}+2\mathrm{i}\log(z^N+C).
\end{gather}
In the $A\to0$ limit we thus recover a single pure point vortex flow (strength $\varGamma_1$), but in the $A\to\infty$ limit we recover the centered-polygon pure point vortex equilibria studied by \citet{Morikawa}.
The stationary polygon consists of a central point vortex with strength $\varGamma_1$ surrounded by $N$ satellite point vortices with strengths $-4\upi$.
Note that, in contrast to the \emph{identical} point vortex limits (up to a translation) of the Stuart vortex solutions \eqref{eq:stuart-sf} as ${A\to0,\infty}$, the polygonal solutions \eqref{eq:sf-poly} have \emph{distinct} pure point vortex limits.

\section{Statement of results: Liouville links and chains}\label{sec:results}
Without any loss of generality, we henceforth set the constants $a$ and $b$ in \eqref{eq:liouville-type} to be $a=1/4\upi$ and $b=-8\upi$, so that $ab=-2$ in \eqref{eq:liouville-sol}.
Next, we rewrite the functions $h'(z)$ and $h(z)$ appearing in the Liouville solution (\ref{eq:liouville-sol}) as
\begin{gather}\label{eq:h'-g'}
h'(z) = A g'(z)
\quad\text{and}\quad
h(z) = A(g(z)+C),
\end{gather}
where $g(z)$ is a primitive of $g'(z)$.
Here we have introduced the complex-analytic function $g'(z)$, the scaling parameter $A>0$, and the complex-valued integration constant $C$.
We will call $g'(z)$ the \emph{input equilibrium function} and $A$ the \emph{hybrid parameter}; these play an important role in the development. 
Substituting \eqref{eq:h'-g'} into \eqref{eq:liouville-sol} we obtain
\begin{gather}\label{eq:hybrid-sf}
\psi(z,\bar{z};A,C) = -\frac{1}{4\upi}\log\left[\frac{A|g'(z)|}{1+A^2|g(z)+C|^2}\right],
\end{gather}
which we call the \emph{hybrid stream function}.

Consider a given stationary point vortex equilibrium of $M$ point vortices located at $z_j$, whose circulations $\varGamma_j$ belong to the set
\begin{gather}\label{eq:circ}
\varGamma_j = -1,\frac{1}{2},1,\frac{3}{2},2,\ldots,
\end{gather}
for $j=1,\ldots,M$. 
From these given point vortex locations and circulations we form the input equilibrium function,
\begin{gather}\label{eq:g'}
g'(z) = \prod_{j=1}^M(z-z_j)^{2\varGamma_j}.
\end{gather}
Further, we identify the subset of positive point vortices in \eqref{eq:g'}, numbering $\widetilde{M}$ and located at $\widetilde{z}_j$ with circulations $\widetilde\varGamma_j$.
The first main result of this paper is that the hybrid stream function \eqref{eq:hybrid-sf} given in terms of the input equilibrium function \eqref{eq:g'}
solves the Liouville-type equation \eqref{eq:liouville-type} and provides a steady solution of the 2D incompressible Euler equation.
The $\widetilde{M}$ positive point vortices in \eqref{eq:g'} remain stationary point vortices in \eqref{eq:hybrid-sf} satisfying the condition \eqref{eq:pv-stationary}, whereas the negative point vortices in \eqref{eq:g'} are smoothed out into the background sea of Liouville-type vorticity.
We refer to the family of hybrid equilibria \eqref{eq:hybrid-sf} as a \emph{Liouville link}.

The second main result of this paper is that the hybrid stream function \eqref{eq:hybrid-sf} interpolates and extrapolates, as a continuous function of the hybrid parameter, between two distinct \emph{pure point vortex equilibria}, i.e. point vortex equilibria with no background vorticity.
These two pure point vortex equilibria have the complex potentials
\begin{gather}\label{eq:limit1}
G(z)=\frac{1}{4\upi\mathrm{i}}\log g'(z)
\quad\text{and}\quad
F(z;C)=\frac{1}{4\upi\mathrm{i}}\log\left[\frac{g'(z)}{(g(z)+C)^2}\right]
\end{gather}
whose imaginary parts are obtained from $\psi(z,\bar{z};A,C)$ {\it viz.} \eqref{eq:hybrid-sf} in the respective limits $A\to 0$ and $A\to\infty$.

The third main result of this paper is that two distinct Liouville links can be sequentially joined together to form a \emph{Liouville chain} after performing a \emph{twist operation} at one end of a Liouville link.
Mathematically, a twist operation is defined as scaling the circulations of the point vortex equilibrium at an end point of the Liouville link, usually the $A\to\infty$ limit, by some \emph{twist parameter} $\alpha$.
Liouville chains can be single-link, $N$-links or infinite-links in length.
Every Liouville link in the chain is an exact solution of the Euler equation given in terms of the \emph{iterated hybrid stream function} (for $n\geq 0$)
\begin{gather}\label{eq:sf-gn}
\psi_n(z,\bar{z};A_n,\boldsymbol{C}_n) 
= -\frac{1}{4\upi}\log\left[\frac{A_n|g_n'(z)|}{1+A_n^2|g_n(z)+C_n|^2}\right].
\end{gather}
The hybrid stream function $\psi_n$ of the $n^\text{th}$ link in the chain depends on one real hybrid parameter ${0<A_n<\infty}$ and $n+1$ complex parameters represented by $\boldsymbol{C}_n=(C_0,\ldots,C_n)$.
Pure point vortex equilibria exist at the end points of every link in the chain and are obtained as limiting cases of \eqref{eq:sf-gn} as $A_n\to0,\infty$.

Note that, in contrast to previous work~\citep{Stuart1967,Crowdy2003a} where a choice for the function $g(z)$ is made, in \eqref{eq:g'} we instead choose its derivative $g'(z)$ in terms of a \emph{known} stationary point vortex equilibrium satisfying \eqref{eq:circ}.
This immediately leads to the question of the existence of a suitable primitive $g(z)$ of $g'(z)$, and seemingly introduces additional complications.
In the first place the velocity field corresponding to the stream function \eqref{eq:hybrid-sf} must be a well-defined, single-valued function so that the stream function represents an actual solution of the 2D Euler equation.
This is in addition to the requirement that the point vortices are stationary according to \eqref{eq:pv-stationary} in order to obtain steady solutions.
These requirements are especially non-trivial to satisfy in the absence of any symmetries.
Our choice satisfies all the above requirements, as shown in \S\ref{sec:theory-links}, and moreover allows us to construct large classes of highly asymmetric equilibria.

We can think of the chain of functions $g_0', g_1', g_2',\ldots$ as an iterated ``transformation'' between point vortex equilibria, a point of view examined in detail by the authors elsewhere \citep{KWCC2020}.
It was actually the considerations concerning hybrid equilibria---of interest in this paper---that motivated that study. 

\section{Constructing a Liouville chain: a detailed example}\label{sec:example}

An example of a non-trivial Liouville link was worked out by the authors in \citet{KWCC2019}.
To illustrate the general theory, in this section we show that this family is one link in an infinite Liouville chain of hybrid equilibria.
The first link in this chain is provided by the special case $N=2$ of the $N$-polygonal equilibria studied by \citet{Crowdy2003a}, the second link is provided by the equilibria studied by \citet{KWCC2019}, and so on.
This Liouville chain of hybrid equilibria is given in terms of the polynomials studied by \citet{Loutsenko2004}.

\subsection{Input seed and the first Liouville link}
We begin by making a very simple choice for the input equilibrium \eqref{eq:g'},
\begin{gather}\label{eq:g'-link1}
	g'(z) = z
	\quad\implies\quad
	g(z) = \frac{z^2}{2},
\end{gather}
corresponding to a single point vortex ($M=1$) at $z_1=0$ with circulation $\varGamma_1=1/2$.
This formula for $g'(z)$ is arrived at by considering the complex potential $f(z)$ given by \eqref{eq:pv-complexpot}, for the single point vortex above, and  taking $g'(z)=\exp(4\upi\mathrm{i} f(z))$.
Substituting \eqref{eq:g'-link1} into \eqref{eq:hybrid-sf}, we obtain the hybrid stream function 
\begin{gather}\label{eq:sf-link1}
	\psi(z,\bar{z};A,C) = -\frac{1}{4\upi}\log\left[\frac{4A|z|}{4+A^2|z^2+2C|^2}\right],
\end{gather}
where the hybrid parameter $A$ varies in the range $0<A<\infty$.
We have not added in an explicit integration constant $C$ to $g(z)$ in \eqref{eq:g'-link1}, this is added separately in \eqref{eq:sf-link1}.
It is clear that the only singularity in \eqref{eq:sf-link1} is a stationary point vortex of strength $+1/2$ at $z=0$, which is seen to be stationary due to a 2-fold rotational symmetry of the solution about the origin.
For all other values of $z$, \eqref{eq:sf-link1} solves \eqref{eq:liouville}.
The stream function and the velocity field are both smooth elsewhere, i.e. the point vortex at $z=0$ is surrounded by everywhere smooth and non-zero Liouville-type vorticity. 
The hybrid stream function \eqref{eq:sf-link1} is therefore a solution of \eqref{eq:liouville-type} with $\widetilde{M}=1$, $\widetilde{z}_1=0$ and $\widetilde\varGamma_1=1/2$.

\begin{figure}
	\centering
	\includegraphics[scale=0.55]{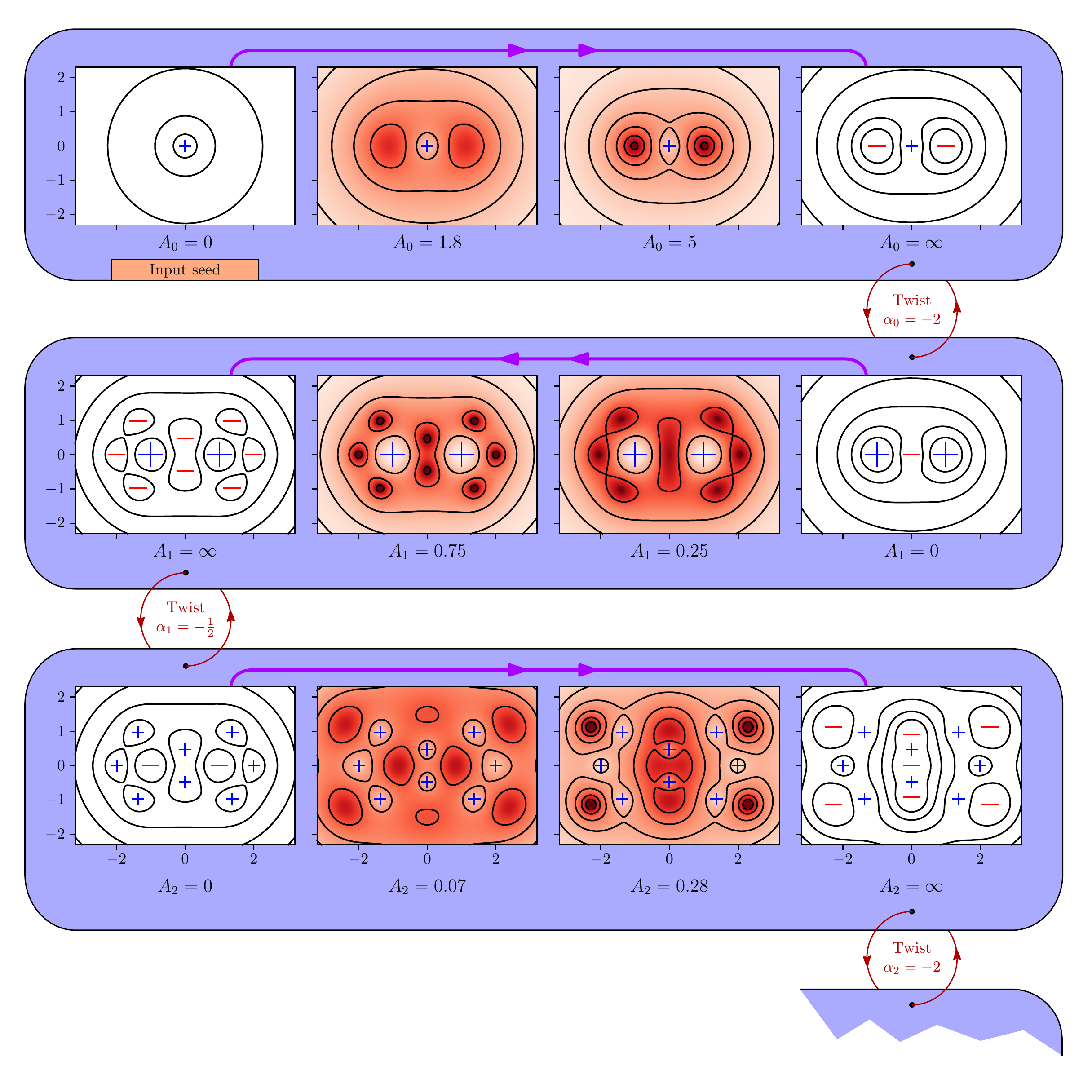}
	\caption{An example of the schematic Liouville chain shown in figure~\ref{fig:schematic}.
		The mathematical details for this example are provided in \S\ref{sec:example}.
		We start with a simple input seed equilibrium, an isolated point vortex in otherwise irrotational flow, with the corresponding $g_0'(z)$ defined by \eqref{eq:g'-link1}. 
		For finite $A_0\neq 0$, the stream function \eqref{eq:sf-link1} is a solution of \eqref{eq:liouville-type} and we refer to this set of hybrid solutions as a Liouville link.
		In the limiting case $A_0=0$ we recover the input seed equilibrium, but for $A_0=\infty$ we obtain a new pure point vortex equilibrium.
		After scaling the circulations of the point vortices in this $A_0=\infty$ equilibrium (we call this a twist operation), we can obtain a new input equilibrium function $g_1'(z)$ allowing us to create a second link in the Liouville chain. 
		The stream function for this second link is given by \eqref{eq:sf-link2}. 
		We can keep adding links to the chain indefinitely, creating a new equilibrium solution with a larger number of vortices at each stage.
		Every pair of point vortex limits are connected by the transformation \eqref{eq:trans-gn} discussed in \citet{KWCC2020}.
		The values of the constants $C_n$ used here are $C_0=-1/2$, $C_1=0$ and $C_2=6$.}
	\label{fig:loutsenko1}
\end{figure}

Rewriting \eqref{eq:sf-link1} as
\begin{gather}
	\psi(z,\bar{z};A,C) = +\frac{1}{4\upi}\log\left[\frac{1}{Az}+A\frac{|z^2+2C|^2}{4z}\right],
\end{gather}
and taking careful limits as $A\to 0,\infty$ leads us to the stream functions
\begin{subequations}\label{eq:sf-limits-link1}
\begin{align}
	\lim_{A\to 0}\left[\psi+\frac{1}{4\upi}\log A\right] 
	&= -\frac{1}{4\upi}\log\left|z\right|,\\
	\lim_{A\to\infty}\left[\psi-\frac{1}{4\upi}\log A\right] 
	&= -\frac{1}{4\upi}\log\frac{4|z|}{|z^2+2C|^2}.
\end{align}
\end{subequations}
These stream functions are clearly the imaginary parts of the complex potentials 
\begin{gather}\label{eq:limits-link1}
	G(z) = \frac{1}{4\upi\mathrm{i}}\log z
	\quad\text{and}\quad
	F(z;C) = \frac{1}{4\upi\mathrm{i}}\log\frac{4z}{(z^2+2C)^2}.
\end{gather}        
The complex potential $G(z)$ corresponds to a single point vortex at $z_1$ with strength $\varGamma_1$; we have thus recovered the input point vortex equilibrium in the limit $A\to 0$.
The other limit $A\to\infty$ yields $F(z;C)$, which corresponds to a colinear three point vortex equilibrium comprising a central point vortex at the origin of circulation $1/2$ and two satellite vortices at $\pm \sqrt{-2C}$ of circulation $-1$ each.
It is a simple matter to check that these three point vortices satisfy \eqref{eq:pv-stat} and are therefore in stationary equilibrium.
Note that \eqref{eq:g'-link1} corresponds to the choice for $g(z)$ made by \citet{Crowdy2003a} with $N=2$.
The family of hybrid equilibria and its point vortex limits are shown as the first Liouville link in figure \ref{fig:loutsenko1}.

\subsection{The necessity of the twist   parameter}
A natural question now arises. 
The above construction started with a known point vortex equilibrium and produced another one.
Can  the same process be re-initiated with a new input equilibrium function in \eqref{eq:hybrid-sf}, corresponding to the new point vortex equilibrium $F(z;C)$ given by \eqref{eq:limits-link1}? 
And will an analogous $A \to \infty$ limit of this hybrid stream function produce {yet another} distinct point vortex equilibrium?

Let us relabel the input equilibrium function in \eqref{eq:g'-link1} as $g_0'$, whose primitive is now called $g_0$.
The corresponding family of hybrid equilibria \eqref{eq:sf-link1} gets relabeled as $\psi_0$ with hybrid parameter $A_0$ and integration constant $C_0$.
Finally, we introduce the notation $G_0$ and $F_0$ for the complex potentials \eqref{eq:limits-link1}.

Instead of the input equilibrium function \eqref{eq:g'-link1}, we now reinterpret the equilibrium $F_0$ as an input, i.e. we choose
\begin{equation}\label{eq:g'-link12}
	g_1'(z) = \frac{z}{(z^2/2 +C_0)^2}
	\quad\implies\quad
	g_1(z) = - {1 \over (z^2/2 +C_0)},
\end{equation}
which is obtained from \eqref{eq:limits-link1} via the formula $g_1' = \exp(4\upi\mathrm{i}F_0)$.
The hybrid stream function follows from substituting \eqref{eq:g'-link12} into \eqref{eq:hybrid-sf}.
After some algebra, we get
\begin{gather}\label{eq:sf-link12}
	\psi_1(z,\bar{z};A_1,C_0,C_1) = -\frac{1}{4\upi}\log\left[
	\frac{4A_1|z|}{|z^2+2C|^2+A_1^2|C_1|^2|z^2+2(C_0-1/C_1)|^2}\right],
\end{gather}
where we have called the new integration constant $C_1$ and the hybrid parameter $A_1$.
Note that \eqref{eq:sf-link12} depends on both integration constants $C_0$ and $C_1$.
Clearly, the only singularity in \eqref{eq:sf-link12} is a point vortex at $z_1=0$ with strength $\varGamma_1=1/2$, and it is stationary due to symmetry.
The hybrid stream function \eqref{eq:sf-link12} is therefore a solution of \eqref{eq:liouville-type} with $\widetilde{M}=1$, $\widetilde{z}_1=0$ and $\widetilde\varGamma_1=1/2$.

Taking limits as $A_1\to 0,\infty$ of \eqref{eq:sf-link12}, in a similar manner as \eqref{eq:sf-limits-link1},
we obtain the complex potentials 
\begin{subequations}\label{eq:limits-link12}
\begin{align}
	G_1(z;C_0) &= \frac{1}{4\upi\mathrm{i}}\log\left[\frac{4z}{(z^2 +2C_0)^2}\right],\\
	F_1(z;C_0,C_1) &= \frac{1}{4\upi\mathrm{i}}\log\left[{1 \over C_1^2}{4z \over (z^2+2(C_0-1/C_1))^2}\right].
\end{align}
\end{subequations}
It is seen from \eqref{eq:limits-link1} and \eqref{eq:limits-link12} that $G_1=F_0$.
We also see that $F_1$ is just a rescaling of $G_1$ in \eqref{eq:limits-link12}.
The point vortex limits of the hybrid stream function \eqref{eq:sf-link12} are not distinct, but the same.
Unfortunately, the choice \eqref{eq:g'-link12} gives nothing new; no new point vortex equilibrium---and hence no new input equilibrium function to continue the iteration---is produced here.

\subsection{Second Liouville link: from three to ten point vortices}
In spite of this apparent setback progress can still be made, and this is where the idea of a twist, using a twist parameter $\alpha$, comes in.
A trivial but crucial observation is: if $f(z)$ is the complex potential for a stationary point vortex equilibrium then so too is $\alpha f(z)$, for any real parameter $\alpha$.
Suppose we take $\alpha_0 = -2$ and rescale $F_0$ in \eqref{eq:limits-link1}, i.e. define
\begin{gather}\label{eq:f-infty}
	G_1(z;C_0) = \alpha_0 F_0(z;C_0) 
	= \frac{1}{4\upi\mathrm{i}}\log\frac{(z^2+2C_0)^4}{z^2},
\end{gather}
where for convenience we have dropped a constant $-(\log 4)/4\upi\mathrm{i}$ from the complex potential.
The complex potential $G_1$ is the previous $F_0$ but now multiplied (we say ``twisted'') by $\alpha_0 =-2$.
All we have done with this twist parameter $\alpha_0$ is to rescale the point vortex circulations without changing the state of hydrodynamic equilibrium.

We now reinitiate the construction, not with the input equilibrium \eqref{eq:g'-link12}, but with the choice $g_1'=\exp(4\upi\mathrm{i} G_1)$ i.e.
\begin{subequations}\label{eq:g'-link2}
\begin{align}
	g_1'(z) &= {(z^2 + 2C_0)^4 \over z^2}, \\
	\implies\quad
	g_1(z) &= {z^7 \over 7} + \frac{8}{5}C_0 z^5 + 8 C_0^2 z^3+ 32 C_0^3 {z} - {16 C_0^4 \over z}.
\end{align}
\end{subequations}
Dropping the constant in \eqref{eq:f-infty} has meant that both the numerator and denominator polynomials in $g_1'(z)$ are monic.
The hybrid stream function \eqref{eq:hybrid-sf} takes the form 
\begin{gather}\label{eq:sf-link2}
	\psi_1(z,\bar{z};A_1,C_0,C_1) = -\frac{1}{4\upi}\log\left[\frac{A_1|z^2+2C_0|^4}
	{|z|^2+A_1^2|z(g_1(z)+C_1)|^2}\right].
\end{gather}
For $A_1>0$, \eqref{eq:sf-link2} is now a solution of  (\ref{eq:liouville-type}) in the case $\widetilde M=2$ with $\widetilde z_1= - \widetilde z_2 =  \sqrt{-2C_0}$ and $\widetilde\varGamma_1=\widetilde\varGamma_2=+2$.
There are now \emph{two} point vortices embedded in the smooth background sea of Liouville-type vorticity.
It is necessary that these point vortices are stationary according to \eqref{eq:pv-stationary} in order to obtain steady solutions. 
This was directly shown to be true for the stream function \eqref{eq:sf-link2} in \citet{KWCC2019} with $C_0=-1/2$ (of course, this also follows from the general theory in \S\ref{sec:theory-links}).

Turning now to the limits $A_1\to 0,\infty$ of \eqref{eq:sf-link2}, we obtain the complex potentials
\begin{subequations}\label{eq:limits-link2}
\begin{align}
	G_1(z;C_0) &= \frac{1}{4\upi\mathrm{i}}\log\frac{(z^2+2C_0)^4}{z^2}, \\
	F_1(z;C_0,C_1) &= \frac{1}{4\upi\mathrm{i}}\log\frac{(z^2+2C_0)^4}{(z(g_1(z)+C_1))^2}.
\end{align}
\end{subequations}
We have obtained the input equilibrium $G_1$ in the $A_1\to 0$ limit, but $F_1$, obtained in the $A_1\to\infty$ limit, is a new pure point vortex equilibrium.
Comparing \eqref{eq:limits-link12} and \eqref{eq:limits-link2}, we see that the twist operation has resulted in a new point vortex equilibrium given by $F_1$ in \eqref{eq:limits-link2}.
This emergent equilibrium is found to comprise two point vortices located at $\pm\sqrt{-2C_0}$ and of circulations $+2$ each, along with eight point vortices located at the roots of the degree-eight polynomial $z(g_1(z)+C_1)$ and of circulations $-1$ each.
With $C_0=-1/2$ the functions in \eqref{eq:g'-link2} are essentially those given in equations (3.8b) and (3.9b) of \cite{KWCC2019}, who explore in detail this set of hybrid equilibria for $0<A_1<\infty$, including the highly non-trivial point vortex equilibrium that emerges in the $A_1 \to \infty$ limit. 
This family of hybrid equilibria and its point vortex limits are shown as the second Liouville link in figure \ref{fig:loutsenko1}.

From this explicit example it should be clear how a function $g'(z)$ associated with a known point vortex equilibrium gives, on substitution into \eqref{eq:hybrid-sf}, a family of hybrid equilibria for any $0 < A < \infty$, called a Liouville link.
These hybrid equilibria are bracketed by two point vortex equilibria corresponding to $A=0$ and $A=\infty$. 
After a suitable twist operation, a second Liouville link can be added to the first, and the procedure can be iterated to produce a Liouville chain.
Figure \ref{fig:loutsenko1} shows the third Liouville link in this chain.
We refer the reader back to figure \ref{fig:schematic} where this process is depicted schematically. 
The example discussed in this section can be continued indefinitely to form an infinite Liouville chain, see \S\ref{ss:loutsenko}.

Not all Liouville chains can be continued indefinitely. 
We give examples of single-link Liouville chains in \S\ref{sec:single}, Liouville chains of finite length in \S\ref{sec:N} and two further examples of infinite Liouville chains in \S\ref{sec:infinite}.

\section{General theory of Liouville links}\label{sec:theory-links}
With our choices for $a$ and $b$ ($a=1/4\upi, b=-8\upi$), the stream function \eqref{eq:liouville-sol} becomes
\begin{gather}\label{eq:liouville-sol1}
\psi(z,\bar{z}) = -\frac{1}{8\upi}\log\left[\frac{|h'(z)|^2}{(1+|h(z)|^2)^2}\right].
\end{gather}
The relations (\ref{eq:def-sf}) can be combined to obtain the expression $u-\mathrm{i}v=2\mathrm{i}\,\partial\psi/\partial z$ for the complex velocity field $u-\mathrm{i}v$ in terms of the stream function.
On using (\ref{eq:liouville-sol1}) for the stream function, we find that the velocity field associated with the Liouville-type vorticity is,
in terms of the analytic function $h(z)$,
\begin{gather}\label{eq:stuart-vel}
u-\mathrm{i} v 
= 2\mathrm{i}\frac{\partial\psi}{\partial z} =\frac{1}{4\upi\mathrm{i}}\left[\frac{h''(z)}{h'(z)}-\frac{2h'(z)\overline{h(z)}}{1+|h(z)|^2}\right].
\end{gather}

The hybrid equilibria constructed in this paper are solutions of the Liouville-type equation \eqref{eq:liouville-type}.
The hybrid vorticity consists of $\widetilde{M}$ point vortices, located at $\widetilde{z}_j$ and with circulations $\widetilde\Gamma_j$, embedded in a sea of Liouville-type vorticity.
Away from the point vortices, i.e. for $z\neq\widetilde{z}_j$, \eqref{eq:liouville-type} is solved by the stream function \eqref{eq:liouville-sol}.
In order to obtain steady solutions of the Euler equation it is necessary and sufficient that the point vortices are stationary.
This is equivalent to a force-free condition on the point vortices.
Similar to \eqref{eq:pv-nsi-vel}, the velocity of a point vortex embedded in a sea of background vorticity can be obtained by considering the non-self-induced part of the velocity field at the point vortex location.
The velocity field $u-\mathrm{i}v$ is given in this case by \eqref{eq:stuart-vel} and the velocity of a point vortex at $\widetilde{z}_k$, for $k=1,2,\ldots,\widetilde{M}$, is ~\citep{LlewellynSmith2011}
\begin{gather}\label{eq:stuart-nsi-vel}
\overline{\frac{\mathrm{d}\widetilde{z}_k}{\mathrm{d}t}} =
\left[(u-\mathrm{i}v)-\frac{\widetilde\Gamma_k}{2\upi\mathrm{i}}\frac{1}{z-\widetilde{z}_k}\right]\Bigg|_{z=\widetilde{z}_k}.
\end{gather}
Then, in order that the point vortex is stationary, we require the local expansion of the velocity field to be without a constant term: it must be of the form \eqref{eq:pv-stationary}.
Note that the leading order term in the regular part of the velocity field in \eqref{eq:pv-stationary} is $O(|z-\widetilde{z}_k|)$ and not necessarily $O(z-\widetilde{z}_k)$ as this is a rotational velocity field.

The central idea of the present paper is to choose the arbitrary function $h(z)$ in the Liouville solution \eqref{eq:liouville-sol1} in terms of a stationary point vortex equilibrium.
More precisely, we choose the function $h'(z)$ as follows.
Given $M$ (note that $M$ is different from $\widetilde{M}$) point vortices in stationary equilibrium at locations $z_j$, with circulations $\varGamma_j$, define the input equilibrium function $g'(z)$ in terms of the point vortex complex potential $f(z)$ {\it viz.} \eqref{eq:pv-complexpot} as
\begin{gather}\label{eq:g'-repeat}
g'(z) = [\exp(2\upi\mathrm{i}\,f(z))]^2 = \prod_{j=1}^M(z-z_j)^{2\varGamma_j}.
\end{gather}
This is the general expression for $g'(z)$ given in \eqref{eq:g'}.
The first observation we make from \eqref{eq:g'-repeat} is that adding a constant to $f(z)$ is equivalent to multiplying $g'(z)$ by a related constant.
We therefore introduce a real parameter $A$ and take the function $h'(z)$---and hence $h(z)$---to be of the form \eqref{eq:h'-g'},
\begin{gather}\label{eq:h'-general}
h'(z) = Ag'(z) = A\prod_{j=1}^M(z-z_j)^{2\varGamma_j}.
\end{gather}
We show that if the point vortex circulations belong to the set \eqref{eq:circ}, then the rational function $h'(z)$ integrates to another rational function, and hence the resulting velocity field \eqref{eq:stuart-vel} is single-valued.
Then, using local expansions in \eqref{eq:stuart-vel} near the zeros and poles of $h(z)$ and $h'(z)$, we show that the vorticity is of the form \eqref{eq:liouville-type} and all the point vortices are stationary according to the condition \eqref{eq:pv-stationary}. 
Among the $M$ point vortices present in the `input' $h'(z)$, $\widetilde{M}$ remain in the hybrid `output' solution; these are the vortices with positive circulation.
The rest of the point vortices (with circulations $-1$) are smoothed out into the background sea of Liouville-type vorticity.

\subsection{Proof that $h(z)$ is rational}\label{ss:proof1}
We begin by showing that restricting the point vortex strengths according to \eqref{eq:circ} leads to a function $h(z)$ that is free of logarithms and hence to a single-valued velocity field \eqref{eq:stuart-vel}.
This proof can also be found in the context of the transformation described in \cite{KWCC2020} but we include it here for completeness.
This condition is equivalent to $h'(z)$ having zero residue at each of its poles which are clearly at point vortex locations $z_k$ with negative circulations $\varGamma_k = -1$.
Near any such $z_k$, we rewrite 
\begin{gather}\label{eq:h'Hk}
h'(z) = A\frac{H_k(z)}{(z-z_k)^{2}},
\end{gather}
where we have defined the functions
\begin{gather}\label{eq:def-Hk}
H_k(z) = \prod_{\substack{j=1\\ j\neq k}}^M (z-z_j)^{2\varGamma_j}
\quad\text{for }k=1,2,\ldots,M.
\end{gather}
Since the vortex positions are non-overlapping, $H_k(z_k)$ is finite and non-zero. 
The series representation for $h'(z)$ near $z_k$ is
\begin{gather}\label{eq:series-ht'}
h'(z) 
= A \left(\frac{H_k(z_k)}{(z-z_k)^{2}} 
+ \frac{H_k^{\,\prime}(z_k)}{(z-z_k)} 
+ \frac{H_k''(z_k)}{2} 
+ \cdots\right).
\end{gather}
Hence $h'(z)$ will have zero residue at $z_k$ if and only if the coefficient $H_k^{\,\prime}(z_k)$ vanishes.
Combining \eqref{eq:def-Hk} and \eqref{eq:pv-stat} yields
\begin{gather}\label{eq:Hk'0}
\frac{H_k^{\,\prime}(z_k)}{H_k(z_k)} 
= \left(\log H_k(z)\right)'\bigg|_{z=z_k} = 
\sum_{\substack{j=1\\ j\neq k}}^M
\frac{\varGamma_j}{z_k-z_j}=0,
\end{gather}
and hence $H_k^{\,\prime}(z_k)=0$ as desired.
Similar arguments show that allowing for $\varGamma_k = -1/2$ in \eqref{eq:circ} would \emph{always} lead to non-rational $h(z)$.
Allowing for larger negative circulations, say $\varGamma_k = -\frac 32$, would require the corresponding coefficient $H_k''(z_k)$ to vanish, which is not true in general.
On the other hand it can happen in specific examples, for instance the trivial example of a single point vortex.

\subsection{Proof that singularities are stationary point vortices}\label{ss:proof2}
With the choice \eqref{eq:h'-general} for $h'(z)$, the stream function is smooth away from the zeros and poles of $h'(z)$ and $h(z)$ and therefore satisfies the modified Liouville equation \eqref{eq:liouville-type}.
Since $h'(z)$ and $h(z)$ are both rational functions, singularities $\widetilde{z}_k$ of the stream function \eqref{eq:liouville-sol1} can only appear at their roots and poles.
It remains to show that at each of these singularities the velocity field is of the form \eqref{eq:pv-stationary} and hence that the stream function satisfies \eqref{eq:liouville-type}.
We have the following three cases to consider:
(a) zeros of $h'(z)$,
(b) poles of $h'(z)$, and
(c) zeros of $h(z)$.
The poles of $h(z)$ and $h''(z)$ coincide with the poles of $h'(z)$, and so do not need to be checked separately. 

Using \eqref{eq:h'-general} we see that the term
\begin{gather}\label{eq:first-term1}
\frac{h''(z)}{h'(z)} = \left(\log h'(z)\right)^\prime = 2\sum_{j=1}^M\frac{\varGamma_j}{z-z_j}
\end{gather}
is proportional to the pure point vortex velocity field \eqref{eq:pv-vel}.
Since the point vortices are stationary, from \eqref{eq:pv-stat} we have,
\begin{gather}\label{eq:first-term2}
\frac{h''(z)}{h'(z)} = \frac{2\varGamma_k}{z-z_k} + O(z-z_k)\quad\text{as }z\to z_k
\end{gather}
near any zero or pole $z_k$ of $h'(z)$.

First consider (a) a zero $z_k$ of $h'(z)$ which is not also a zero of $h(z)$. 
The second term in \eqref{eq:stuart-vel} vanishes at $z_k$ and we have from \eqref{eq:first-term2}
\begin{gather}
u-\mathrm{i}v = \frac{1}{2\upi\mathrm{i}}\frac{\varGamma_k}{z-z_k} + O(z-z_k)\quad\text{as }z\to z_k,
\end{gather}
which corresponds to a stationary point vortex at $z_k$ with circulation $\widetilde\Gamma_k=\varGamma_k$.
Note that the zeros of $h'(z)$ correspond to point vortices with positive circulations $\varGamma_k$.
Next, considering (b), it was already noted by \cite{Crowdy2003a} that a simple-pole term in $h(z)$ leads to a smooth velocity field at the location of the pole. 
Since by \eqref{eq:circ} the only poles of $h'(z)$ are second-order poles, the only poles of $h(z)$ will be simple poles. 
Going back to the stream function \eqref{eq:liouville-sol1} we see that at a simple pole $z_k$ of $h(z)$, the argument of the logarithm has a constant non-zero leading term, and hence the stream function is regular at $z_k$ due to the structure of the Liouville solution.
The same conclusion may also be reached by expanding \eqref{eq:stuart-vel} near $z_k$.

It remains to consider (c), of which there are two types, simple and multiple zeros.
At a simple zero $\widehat{z}_k$ (say) of $h(z)$ the second term in \eqref{eq:stuart-vel} vanishes while the first term is regular since $h'(z)$ is regular at $\widehat{z}_k$. 
The velocity field is therefore regular at a simple zero of $h(z)$.
If $\widehat{z}_j$ is a multiple zero of $h(z)$ then it must also be a zero of $h'(z)$ and so $\widehat{z}_j=z_k$ for some $k=1,\ldots,M$.
The second term in \eqref{eq:stuart-vel} is again zero but the first term contributes a pole due to the zero of $h'(z)$.
Since $z_k$ is a root of $h'(z)$ with multiplicity $2\varGamma_k$ by construction (see \eqref{eq:h'-general}), it must be a root of $h(z)$ with multiplicity $2\varGamma_k + 1$.
Thus we can write
\begin{equation}
\label{eq:Gk}
h(z) = (z-z_k)^{2\varGamma_k+1} G_k(z) 
\end{equation}
for some rational function $G_k(z)$ with $G_k(z_k) \ne 0$. 
Differentiating \eqref{eq:Gk} yields
\begin{align}
\label{eq:Gk'}
h'(z) = (z-z_k)^{2\varGamma_k} \big((2\varGamma_k+1)G_k(z) + (z-z_k)G_k'(z)\big),
\end{align}
and hence that $G_k(z)$ is related to the function $H_k(z)$ defined in \eqref{eq:def-Hk} via
\begin{align}
\label{eq:diffme}
(2\varGamma_k+1)G_k(z) + (z-z_k)G_k'(z) = H_k(z).
\end{align}
Differentiating \eqref{eq:diffme} and substituting $z=z_k$, we find 
\begin{equation}
(2\varGamma_k+2)G_k'(z_k) = H_k'(z_k) = 0,
\end{equation}
where the last equality follows from \eqref{eq:pv-stat} exactly as in \eqref{eq:Hk'0}.
In particular, since ${\varGamma_k \ne -1}$, we deduce that $G_k'(z_k) = 0$.
Substituting in \eqref{eq:Gk'} and differentiating once more we calculate the local expansions for $h'(z)$ and $h''(z)$ near $z_k$ to be
\begin{subequations}
	\begin{align}
	h'(z) &= (z-z_k)^{2\varGamma_k}\left((2\varGamma_k+1)G_k(z_k)+O(z-z_k)^2\right), \\
	h''(z) &= (z-z_k)^{2\varGamma_k-1}\left(2\varGamma_k(2\varGamma_k+1)G_k(z_k)+O(z-z_k)^2\right).
	\end{align}
\end{subequations}
This leads to the velocity field \eqref{eq:stuart-vel}
\begin{gather}
u-\mathrm{i}v = \frac{1}{2\upi\mathrm{i}}\frac{\varGamma_k}{z-z_k} + O(z-z_k)\quad\text{as }z\to z_k,
\end{gather}
which corresponds to a stationary point vortex at $z_k$ of strength $\widetilde\Gamma_k=\varGamma_k$.

To summarise: Under the restriction \eqref{eq:circ}, a rational $h'(z)$ defined by \eqref{eq:h'-general} leads to a rational $h(z)$.
The roots of $h'(z)$, which correspond to the positive circulation point vortices in the input equilibrium \eqref{eq:h'-general}, are preserved as point vortices at the same locations (renamed as $\widetilde{z}_k$) with their strengths remaining the same.
The poles of $h'(z)$, which correspond to point vortices with circulations $-1$, become the background sea of smooth Liouville-type vorticity.

If the value of the constant $b$ is chosen to be different from $-8\upi$, all elements of the proofs above remain the same except for the strengths $\widetilde\varGamma_k$, which now scale as $\widetilde\varGamma_k=(-8\upi/b)\varGamma_k$.
Appendix \ref{app:convention} contains more details.

\section{Liouville chains}\label{sec:theory-chains}

\subsection{Point vortex limits of a Liouville link and a transformation between them}\label{ss:limits}
The stream function \eqref{eq:hybrid-sf} for the Liouville link contains a real parameter $A$ and a complex parameter $C$. 
In this section, we are interested in the behavior of the hybrid stream function in the limits $A\to 0,\infty$ and show that in these limits the rotational solutions approach distinct stationary point vortex equilibria. 

Taking $A>0$ and rearranging the argument of the logarithm in the hybrid stream function \eqref{eq:hybrid-sf} we obtain
\begin{subequations}
	\begin{align}
	\psi(z,\bar{z};A,C)
	&= +\frac{1}{4\upi}\log\left[\frac{1+A^2|g(z)+C|^2}{A|g'(z)|}\right] \\
	&= +\frac{1}{4\upi}\log\left[\frac{1}{A|g'(z)|} + A\frac{|g(z)+C|^2}{|g'(z)|}\right]. 
	\label{eq:sf-limit}
	\end{align}
\end{subequations}
Note that we can take $A>0$ without loss of generality since if $A$ were negative or indeed complex, then $|A|>0$ would appear in \eqref{eq:sf-limit} in place of $A$. 

Consider now the limiting case $A\to 0$. 
In this case the second term in the argument of the logarithm drops out and the stream function \eqref{eq:sf-limit} can be written, after renormalising for the infinite term $-(1/4\upi)\log A$, as 
\begin{subequations}
	\begin{align}\label{eq:sf-Ato0}
	\lim_{A\to 0}\left[\psi(z,\bar{z};A,C)+\frac{1}{4\upi}\log A\right] 
	&= -\frac{1}{4\upi} \log |g'(z)|.
	\intertext{Similarly, in the limiting case $A\to\infty$ we see that the first term in the argument of the logarithm drops out and the stream function \eqref{eq:sf-limit} can be written, after renormalising for the infinite term $(1/4\upi)\log A$, as}
	\label{eq:sf-Atoinf}
	\lim_{A\to\infty}\left[\psi(z,\bar{z};A,C)-\frac{1}{4\upi}\log A\right] 
	&= -\frac{1}{4\upi}\log \frac{|g'(z)|}{|g(z)+C|^2}.
	\end{align}
\end{subequations}

The stream functions \eqref{eq:sf-Ato0} and \eqref{eq:sf-Atoinf} are respectively the imaginary parts of the complex potentials
\begin{gather}\label{eq:limit2}
G(z)=\frac{1}{4\upi\mathrm{i}}\log g'(z)
\quad\text{and}\quad
F(z;C)=\frac{1}{4\upi\mathrm{i}}\log\left[\frac{g'(z)}{(g(z)+C)^2}\right],
\end{gather}
determined by the input equilibrium $g'(z)$. 
Comparing \eqref{eq:limit2} with \eqref{eq:g'-repeat} we see that $G(z)=f(z)$ is the complex potential of the stationary point vortex equilibrium  we started out with.
The complex potential $F(z;C)$ also corresponds to a stationary point vortex equilibrium, distinct from $G(z)$, as explained below. 

In order to relate the two different complex potentials $G(z)$ and $F(z)$ defined in \eqref{eq:limit2}, we first define a new function $\widehat{g}'(z)$ via the \emph{transformation} 
\begin{gather}\label{eq:trans-g}
	g'(z)\mapsto\widehat{g}'(z) = \widehat{E}\left[\frac{g'(z)}{(g(z)+C)^2}\right]^\alpha,
\end{gather}
where $\alpha$ is the twist parameter and $\widehat{E}$ is some constant.
\cite{KWCC2020} show that the transformation \eqref{eq:trans-g} takes a given stationary point vortex equilibrium $g'(z)$ into a new equilibrium $\widehat{g}'(z)$ if the point vortex circulations in $g'(z)$ belong to the set \eqref{eq:circ}.
The primitive of $g'(z)$ is then also a rational function and $\widehat{g}'(z)$ {\it viz.} \eqref{eq:trans-g} takes the same mathematical form as $g'(z)$ {\it viz.} \eqref{eq:g'}, but with some of the vortex positions and circulations changed. 
The proof for this assertion is very similar to the proof presented in \S\ref{ss:proof1} and \S\ref{ss:proof2}, and is detailed in \cite{KWCC2020}.

The stationary point vortex equilibria connected by the transformation \eqref{eq:trans-g} are closely related to the complex potentials $G(z)$ and $F(z;C)$ in \eqref{eq:limit2} of stationary point vortex equilibria that exist at the end points of a Liouville link.
Indeed, let us define $\widehat{G}(z;C)$ after scaling the circulations in $F(z;C)$ by $\alpha$, and adding a constant in terms of $\widehat{E}$, 
\begin{gather}\label{eq:fhat}
	\widehat{G}(z;C)=\alpha F(z;C) + \frac{1}{4\upi\mathrm{i}}\log\widehat{E}.
\end{gather}
We can rewrite \eqref{eq:limit2} in terms of $\widehat{G}(z;C)$ as 
\begin{gather}\label{eq:limit3}
G(z)=\frac{1}{4\upi\mathrm{i}}\log g'(z)
\quad\text{and}\quad
\widehat{G}(z;C)=\frac{1}{4\upi\mathrm{i}}\log\widehat{g}'(z).
\end{gather}
Thus $G(z)=f(z)$ is the stationary point vortex equilibrium we started out with, defined by $g'(z)$, whereas $\widehat{G}(z;C)$ (or $F(z;C)$) is a new stationary point vortex equilibrium defined by $\widehat{g}'(z)$ which is given by the transformation \eqref{eq:trans-g}.
We emphasize that $\widehat{G}(z;C)$ is obtained after scaling all the circulations in the $A\to\infty$ limit \eqref{eq:limit2} of the Liouville link by the twist parameter $\alpha$.
Finally, we note that by definition $G(z)$ is independent of $C$.

\subsection{Twist operations and Liouville chains }\label{ss:iterated}
A brief summary of Liouville links and their limits that we have discussed so far in \S\ref{sec:theory-links} and \S\ref{ss:limits} is as follows. The input equilibrium function $g'(z)$ in the hybrid stream function \eqref{eq:hybrid-sf} of the Liouville link can be chosen to be of the form \eqref{eq:g'}, consisting of stationary point vortices whose circulations belong to the set \eqref{eq:circ}. 
Then for any finite value of the parameter $A$ the Liouville link solution exists.
The transformation \eqref{eq:trans-g} allows us to jump directly between the end points of the Liouville link which are pure point vortex equilibria.
These latter equilibria are recovered as distinct limits of the rotational hybrid solutions, as $A\to0,\infty$.

The transformation \eqref{eq:trans-g} between stationary pure point vortex equilibria can sometimes be iterated to produce hierarchies of pure point vortex equilibria consisting of increasing numbers of point vortices with each iteration \citep{KWCC2020}. 
It is also possible to have the iteration continue indefinitely.
At a given $n^\text{th}$ stage of the iteration, if $g_n'(z)$ is a stationary point vortex equilibrium of the form \eqref{eq:g'} with the circulations belonging to the set \eqref{eq:circ}, then $g_n(z)$ is a rational function and $g_{n+1}'(z)$ defined by the \emph{iterated transformation} (for some constants $E_n$)
\begin{gather}\label{eq:trans-gn}
	g_{n+1}'(z) = E_{n+1}\left[\frac{g_n'(z)}{(g_n(z)+C_n)^2}\right]^{\alpha_n},
\end{gather}
is a stationary point vortex equilibrium.
A new twist parameter $\alpha_n$ is defined at every stage of the iteration \eqref{eq:trans-gn}.
If $\alpha_n$ can be chosen so that the circulations in $g_{n+1}'(z)$ also belong to the set \eqref{eq:circ}, then the iteration can be continued.

\cite{KWCC2020} show that, if we choose at any stage the special value $\alpha_n=1$, then the resulting equilibrium will simply be a space-shifted version of the equilibrium at the previous stage.
They also discuss various classes of stationary point vortex equilibria that can be generated from a ``seed equilibrium function'' of the form
\begin{gather}\label{eq:general-seed}
g_0'(z)=z^{2\varGamma}
\end{gather} 
with different choices of $\varGamma$ and $\alpha_n$.
Here $\varGamma$ is the circulation of the seed point vortex. 
In this manner the Adler-Moser polynomials found by \citet{Adler1978} and the two polynomial hierarchies discussed by \citet{Loutsenko2004} are all produced from the same seed \eqref{eq:general-seed} through the iterated transformation \eqref{eq:trans-gn}.
The point vortices in equilibrium are at the roots of successive polynomials in these hierarchies.

A Liouville chain is a sequence of Liouville links, joined together by applying a twist operation at an end point of each Liouville link.
Every link in the Liouville chain is a family of hybrid stream functions, defined by \eqref{eq:sf-gn}, and continuously parametrised by $0<A_n<\infty$.
The end points of every Liouville link are pure point vortex equilibria obtained in the limits $A_n\to 0,\infty$ and have the complex potentials
\begin{gather}
	G_n(z) = \frac{1}{4\upi\mathrm{i}}\log g_n'(z)
	\quad\text{and}\quad
	F_n(z) = \frac{1}{4\upi\mathrm{i}}\log\left[\frac{g_n'(z)}{(g_n(z)+C_n)^2}\right].
\end{gather}
The twist operation is encoded by the twist parameter $\alpha_n$ which scales the circulations of all the point vortices at the $A_n\to\infty$ limit of the Liouville link. 
The point vortex equilibrium obtained after scaling the circulations and adding a constant is used to build the next link in the chain:
\begin{gather}
	G_{n+1}(z) = \alpha_n F_n(z) + \frac{1}{4\upi\mathrm{i}}\log E_{n+1}.
\end{gather}
The stream function at the $n^\text{th}$ stage will contain $n+2$ parameters: the real parameter $A_n$ and $n+1$ complex parameters $\boldsymbol{C}_n$.
Every function $g_n'(z)$ in the chain is of the form \eqref{eq:g'} and is obtained via the iterated transformation \eqref{eq:trans-gn}.
As long as the sequence of rational functions $g_n'(z)$ corresponding to stationary pure point vortex equilibria exists, a corresponding sequence of stream functions given by \eqref{eq:sf-gn} also exists.

\section{Single-link Liouville chains}\label{sec:single}

\subsection{Liouville link between three and eleven point vortices }
Consider a stationary equilibrium~\citep{ONeil2006,KWCC2020} consisting of three point vortices of strengths $3$, $3/2$, $-1$ located at $-2$, $1$, $0$. 
The input equilibrium function $g'(z)$ corresponding to this point vortex equilibrium is written down from \eqref{eq:g'}:
\begin{gather}\label{eq:g'-oneil3}
g'(z) = \frac{(z+2)^6(z-1)^3}{z^2}.
\end{gather}
Since the point vortex circulations are restricted to the set \eqref{eq:circ}, it follows that the primitive of $g'(z)$ must be rational.
Indeed, we get after an explicit calculation,
\begin{gather}\label{eq:g-oneil3}
	g(z) = \frac{z^8}{8}+\frac{9 z^7}{7}+\frac{9 z^6}{2}+3 z^5-18 z^4-36 z^3+24 z^2+144 z+\frac{64}{z}.
\end{gather} 
Substituting in \eqref{eq:hybrid-sf} and taking $C$ as an integration constant, we get the hybrid stream function 
\begin{gather}\label{eq:sf-oneil3}
	\psi(z,\bar{z};A,C) = -\frac{1}{4\upi}\log\left[
	\frac{A|z+2|^6|z-1|^3}{|z|^2 + A^2|z(g(z)+C)|^2}\right],
\end{gather}
which defines a Liouville link solution for the range of parameter values $0<A<\infty$.
It is clear from \eqref{eq:sf-oneil3} that the point vortex at $z=0$ in \eqref{eq:g'-oneil3}, with circulation $-1$, has been smoothed out.
The two point vortices at $z=-2$ and $z=1$ in \eqref{eq:g'-oneil3} remain embedded in the flow retaining the values of their circulations, are stationary, and are surrounded by an everywhere smooth rotational flow of Liouville-type.

The transformation \eqref{eq:trans-g} defines the function $\widehat{g}'(z)$. 
Setting $\alpha=1$, $\widehat{E}=1/64$ in \eqref{eq:trans-g} and using \eqref{eq:g'-oneil3}, \eqref{eq:g-oneil3} we get
\begin{gather}\label{eq:ghat'-oneil3}
	\widehat{g}'(z) = \frac{(z+2)^6(z-1)^3}{(8z(g(z)+C)^2},
\end{gather}
where $8z(g(z)+C)$ is a degree-9 monic polynomial as seen from \eqref{eq:g-oneil3}.
Comparing the form of $\widehat{g}'(z)$ with \eqref{eq:g'}, we see that it consists of eleven point vortices: the two positive point vortices at $z=-2$ and $z=1$ together with nine negative point vortices of strength $-1$ located at the roots of the polynomial $8z(g(z)+C)$. 

\begin{figure}
	\centering\includegraphics[scale=0.5]{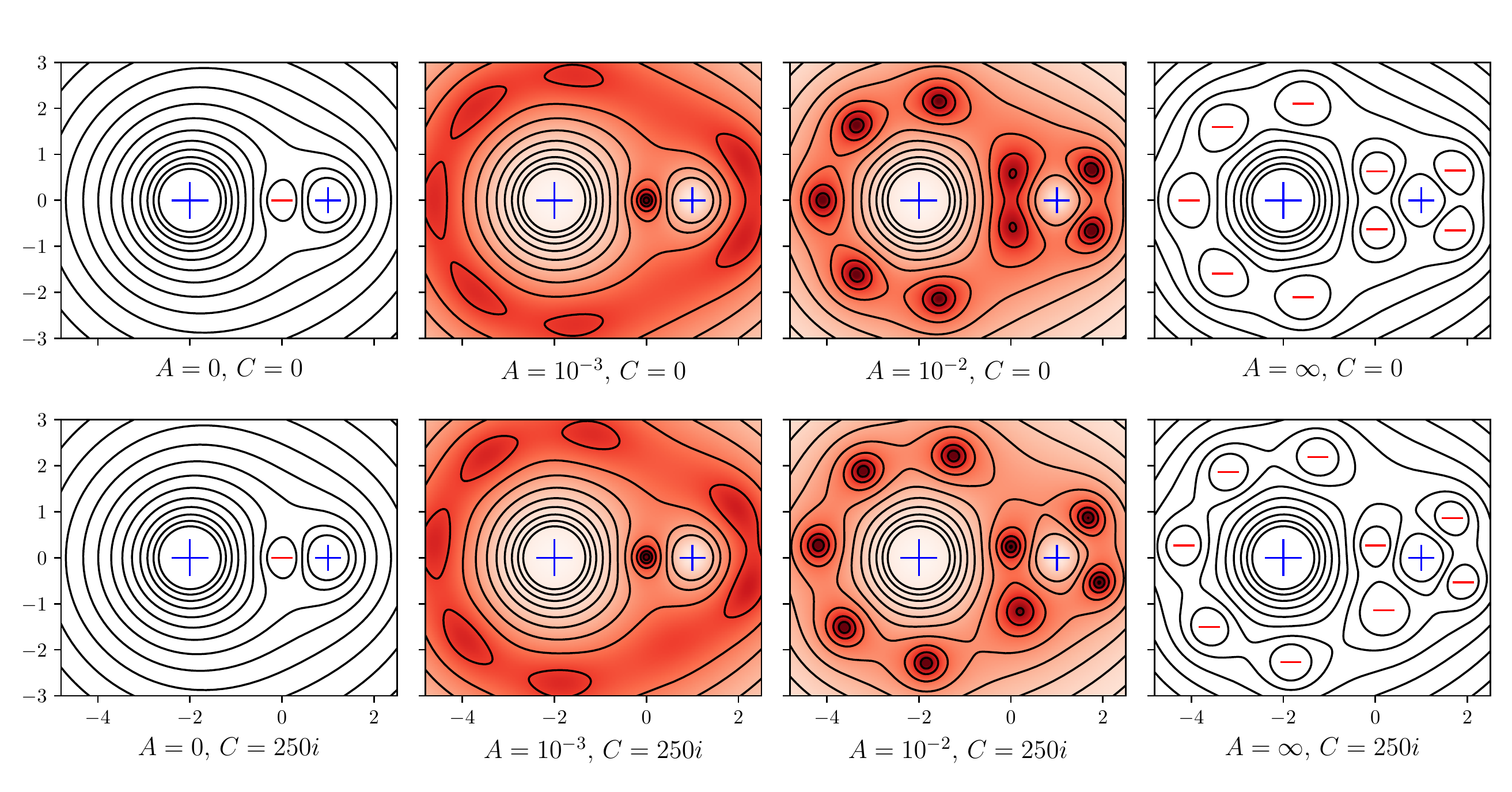}
	\caption{Streamline patterns for the stream function \eqref{eq:hybrid-sf} with the input equilibrium function \eqref{eq:g'-oneil3}.
		The Liouville link exists for the  range of parameter values $0<A<\infty$.
		The formation of the limiting point vortex equilibria \eqref{eq:limit3}, with $g'(z)$ and $\widehat{g}'(z)$ given by \eqref{eq:g'-oneil3} and \eqref{eq:ghat'-oneil3}, can be seen as $A$ becomes small and large.
		This process is shown for two values of $C$. 
		When $C=0$ the hybrid solutions are symmetric with respect to the $x$-axis but when $C=100+180\mathrm{i}$ (complex-valued) this symmetry is lost.
		The $A= 0$ limit is independent of $C$ by definition.}
	\label{fig:oneil3}
	\centering\includegraphics[scale=0.5]{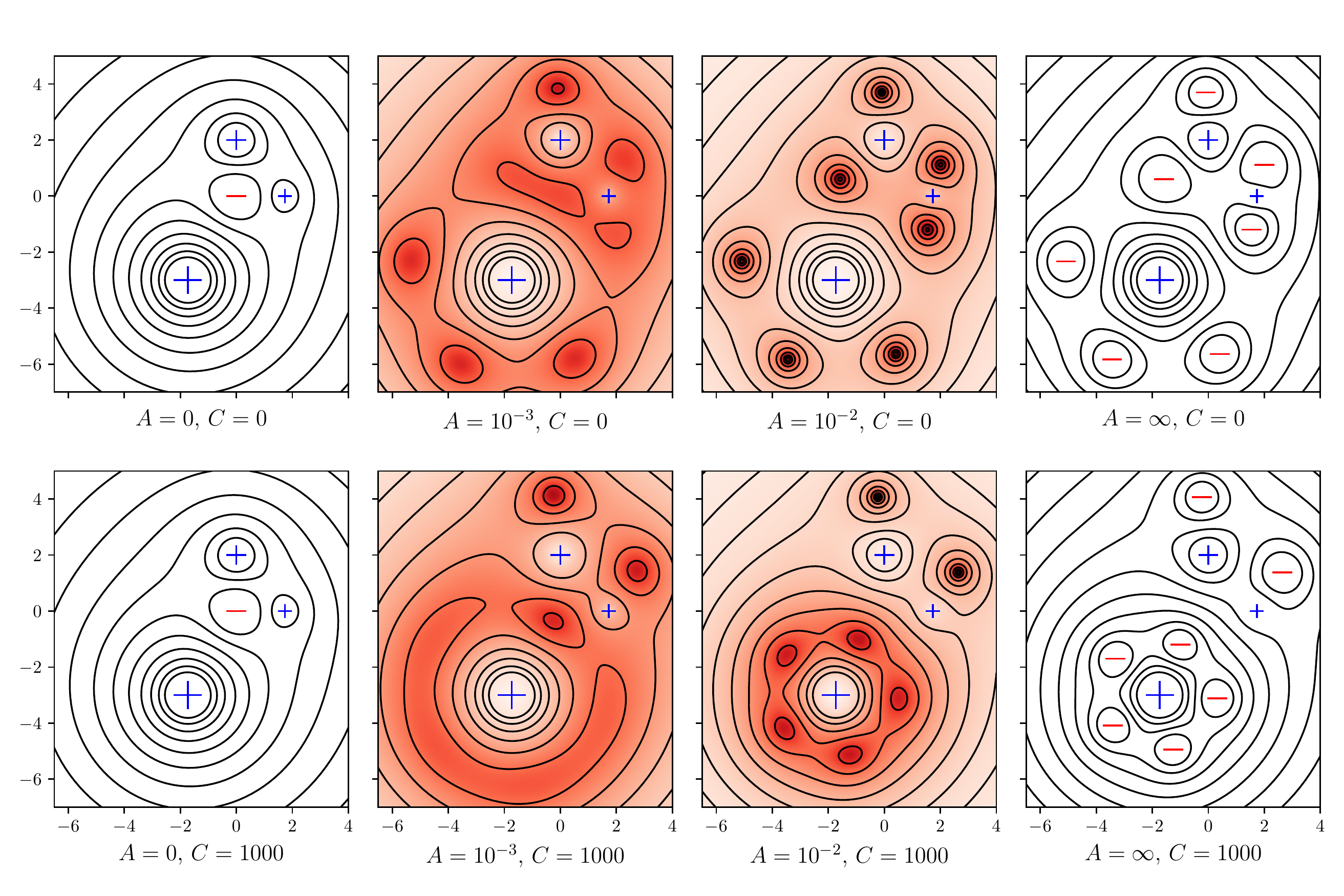}
	\caption{Streamline patterns for the hybrid stream function \eqref{eq:hybrid-sf} with input equilibrium \eqref{eq:g'-oneil4}.
		In the limit as $A\to0,\infty$ the hybrid solution goes over into the stationary point vortex equilibria given by \eqref{eq:limit3}, with $g'(z)$ and $\widehat{g}'(z)$ as in \eqref{eq:g'-oneil4} and \eqref{eq:ghat'-oneil4}.
		Varying the integration constant $C$ alters the locations of the centers and saddles in the flow.}
	\label{fig:oneil4}
\end{figure}

In the limits $A\to 0$ and $A\to\infty$, the stream function \eqref{eq:sf-oneil3} becomes the imaginary part of the complex potentials \eqref{eq:limit3} with $g'(z)$ and $\widehat{g}'(z)$ given by \eqref{eq:g'-oneil3} and \eqref{eq:ghat'-oneil3}.
To see this notice that the stream function \eqref{eq:sf-oneil3} can be re-written as 
\begin{gather}
	\psi(z,\bar{z};A,C) = +\frac{1}{4\upi}\log\left[
	\frac{1}{A}\frac{|z|^2}{|z+2|^6|z-1|^3}+A\frac{|z(g(z)+C)|^2}{|z+2|^6|z-1|^3}\right],
\end{gather}
which in the limits $A\to 0,\infty$ becomes
\begin{subequations}
\begin{align}
	\lim_{A\to 0}\left[\psi+\frac{1}{4\upi}\log A\right] &= -\frac{1}{2\pi}\log\left|\frac{(z+2)^3(z-1)^{3/2}}{z}\right|,\\
	\lim_{A\to\infty}\left[\psi-\frac{1}{4\upi}\log A\right] &= -\frac{1}{2\pi}\log\left|\frac{(z+2)^3(z-1)^{3/2}}{z(g(z)+C)}\right|.
\end{align}
\end{subequations}
The stream function \eqref{eq:sf-oneil3} is thus a Liouville link between the three point vortex equilibrium represented by \eqref{eq:g'-oneil3} and the eleven point vortex equilibrium represented by \eqref{eq:ghat'-oneil3}.
The streamline patterns for this Liouville link are shown in figure~\ref{fig:oneil3} for various values of parameters $A$ and $C$; the figure also shows the limiting point vortex equilibria.

\subsection{Liouville link between four and ten point vortices }
Figure~\ref{fig:oneil4} shows the streamline patterns for the stream function \eqref{eq:hybrid-sf} with the input equilibrium function \eqref{eq:g'} formed from a four point vortex equilibrium~\citep{ONeil2006,KWCC2020}
\begin{gather}\label{eq:g'-oneil4}
	g'(z) = \frac{(z+\sqrt{3}+3\mathrm{i})^4(z-2\mathrm{i})^2(z-\sqrt{3})}{z^2}.
\end{gather}
The point vortex strengths in \eqref{eq:g'-oneil4} satisfy the constraints \eqref{eq:circ} and the rational function $g(z)$ is
\begin{gather}\label{eq:g-oneil4}
\begin{multlined}[0.85\textwidth]
	g(z) = \frac{z^6}{6} + \frac{1}{5} (3\sqrt{3}+8\mathrm{i}) z^5 - (1-3\sqrt{3}\mathrm{i}) z^4 + 4(2\sqrt{3}+3\mathrm{i}) z^3 \\
	- 12(1-3\sqrt{3}\mathrm{i}) z^2 + 24(\sqrt{3}-9\mathrm{i}) z + \frac{288(\sqrt{3}+3\mathrm{i})}{z}.
\end{multlined}
\end{gather}
Setting the twist parameter $\alpha=1$ along with $\widehat{E}=1/36$, the transformed point vortex equilibrium \eqref{eq:trans-g} is given by 
\begin{subequations}\label{eq:ghat'-oneil4}
\begin{gather}
	\widehat{g}'(z) = \frac{(z+\sqrt{3}+3\mathrm{i})^4(z-2\mathrm{i})^2(z-\sqrt{3})}{q^2(z)},
\intertext{where the polynomial $q(z)$ is}
\begin{multlined}[0.85\textwidth]
	q(z) = z^7 + \frac{6}{5} (3\sqrt{3}+8\mathrm{i}) z^6 - 6(1-3\sqrt{3}\mathrm{i}) z^5 + 24(2\sqrt{3}+3\mathrm{i}) z^4 \\
	- 72(1-3\sqrt{3}\mathrm{i}) z^3 + 144(\sqrt{3}-9\mathrm{i}) z^2 + 6Cz + 1728(\sqrt{3}+3\mathrm{i}).
	\end{multlined}
\end{gather}
\end{subequations}

The Liouville link shown in figure~\ref{fig:oneil4} consists of the three positive point vortices in \eqref{eq:g'-oneil4} embedded in a background smooth Liouville-type vorticity for the range of parameter values $0<A<\infty$. 
The $A\to 0,\infty$ limits of the Liouville link are pure point vortex equilibria with four and ten point vortices.
Their complex potentials are given by \eqref{eq:limit3}, where $g'(z)$ and $g(z)$ are given by \eqref{eq:g'-oneil4} and \eqref{eq:ghat'-oneil4}.

\section{$N$-link Liouville chains}\label{sec:N}  

\begin{figure}
	\centering\includegraphics[scale=0.5]{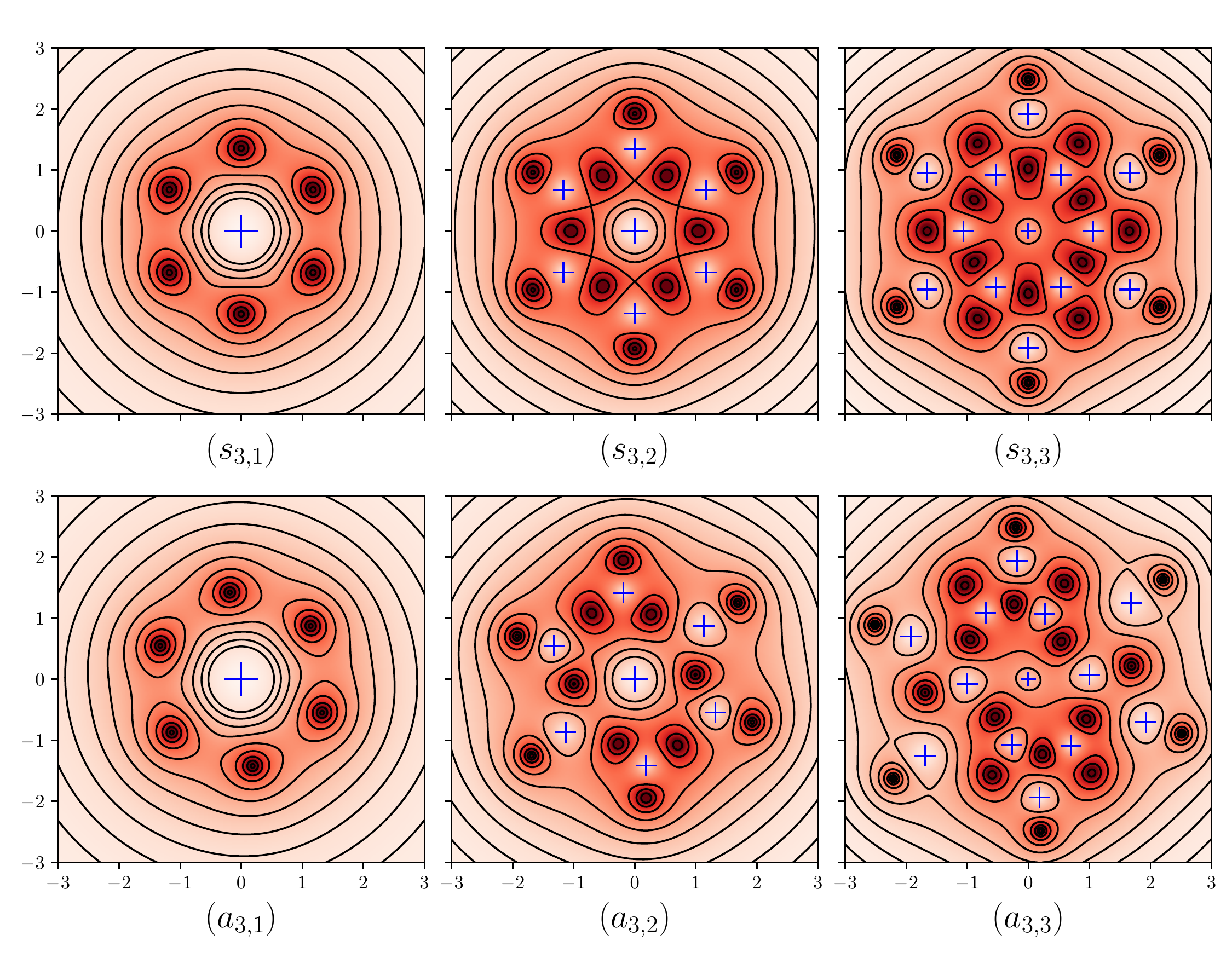}
	\caption{Symmetric $(s_{3,n})$ and asymmetric $(a_{3,n})$ streamline patterns in successive links $n=0,1,2$ of a 3-link Liouville chain.
		The $n^\text{th}$ link is obtained by substituting the rational function $g_n'(z)$ in \eqref{eq:terminating3} into the iterated hybrid stream function \eqref{eq:sf-gn}.
		The circulation of the point vortex at the origin is a positive half-integer which progressively decreases until it reaches $+1/2$, at which point the chain ends. 
		The point vortex limits of these hybrid equilibria are shown as (S3) and (A3) in figure 6 of \cite{KWCC2020}. 
		See table \ref{tab:vals} for the values of $A_n$ and $C_n$ in these plots.}
	\label{fig:terminating}
\end{figure}
We have identified an interesting set of finite-link Liouville chains that terminate after $N\geq 1$ steps. 
If we choose the circulations of the seed equilibria \eqref{eq:general-seed} to be half-integers,
\begin{gather}\label{eq:Gamma-N}
	\varGamma = \frac{2N-1}{2}\quad\text{for}\quad N\geq 1,
\end{gather}
then the corresponding input equilibrium functions
\begin{gather}\label{eq:terminating-seed}
g_0^{\,\prime}(z) = z, z^3, z^5,\ldots,
\end{gather}
together with the twist parameters $\alpha_n=-1$ (for $n\geq 1$), produce precisely $N$-links in the Liouville chain. 

The rational functions for $N=1$, produced by the iterated transformation \eqref{eq:trans-gn}, are
\begin{gather}
g_0'(z) =  z, 
\qquad
g_1'(z) =  \frac{\left(z^{2} + 2 C_{0}\right)^{2}}{z}.
\end{gather}
Here, as in the single-link examples discussed in \S\ref{sec:single}, the constants $E_n$ are chosen so that $g_n'(z)$ have monic numerator and denominator polynomials (see appendix \ref{app:convention}).
The rational function $g_0'(z)$ has a rational primitive $g_0(z)$ since the circulations in $g_0'(z)$ satisfy \eqref{eq:circ}. 
On the other hand, the transformed equilibrium $g_1'(z)$ has a point vortex with circulation $-1/2$ and hence cannot have a rational primitive.
The stream function $\psi_0(z,\bar{z})$ in \eqref{eq:sf-gn} with the input equilibrium function $g_0'(z)$ is therefore a hybrid equilibrium but $\psi_1(z,\bar{z})$ with input equilibrium function $g_1'(z)$ is not.
This is a single-link Liouville chain.

The rational functions for $N=2$ can be obtained from \eqref{eq:trans-gn}:
\begin{gather}
\begin{gathered}
\begin{aligned}
g_0'(z) &=  z^{3},\\
g_1'(z) &=  \frac{\left(z^{4} + 4 C_{0}\right)^{2}}{z^{3}},
\end{aligned}
\qquad 
g_2'(z) = \frac{\left(z^{8} + 24 C_{0} z^{4} + 6 C_{1} z^{2} - 48 C_{0}^{2}\right)^{2}}{z \left(z^{4} + 4 C_{0}\right)^{2}}.
\end{gathered}	
\end{gather}
As before $g_0'(z)$ has a rational primitive and $g_2'(z)$ does not.
The rational function $g_1'(z)$ has a single point vortex at the origin whose circulation is a \emph{negative half-integer}, $-3/2$, while the circulations of its remaining point vortices satisfy \eqref{eq:circ}.
Nevertheless it has a rational primitive $g_1(z)$. 
The stream functions $\psi_0(z,\bar{z})$ and $\psi_1(z,\bar{z})$ in \eqref{eq:sf-gn} are therefore hybrid equilibria while $\psi_2(z,\bar{z})$ is not.
The point vortex at the origin in $g_1'(z)$ remains a point vortex in the hybrid solution $\psi_1(z,\bar{z})$ but with circulation $+1/2$, as can be checked with a local expansion.
This is a 2-link Liouville chain.

The rational functions in the case $N=3$ are
\begin{subequations}\label{eq:terminating3}
\begin{gather}
\begin{alignedat}{2}
	g_0'(z) &= z^{5},
	& g_2'(z) &= \frac{\left(z^{12} + 48 C_{0} z^{6} + 8 C_{1} z^{4} - 72 C_{0}^{2}\right)^{2}}{z^{3} \left(z^{6} + 6 C_{0}\right)^{2}}, \\
	g_1'(z) &= \frac{\left(z^{6} + 6 C_{0}\right)^{2}}{z^{5}},\qquad
	& g_3'(z) &= \frac{q^2(z)}{z \left(z^{12} + 48 C_{0} z^{6} + 8 C_{1} z^{4} - 72 C_{0}^{2}\right)^{2}},
\end{alignedat}
\end{gather}		
where the numerator $q(z)$ of $g_3'(z)$ is
\begin{gather}
\begin{multlined}[0.75\textwidth]
	p(z) = z^{18} + 216 C_{0} z^{12} + 80 C_{1} z^{10} + 10 C_{2} z^{8} - 4320 C_{0}^{2} z^{6} - 960 C_{0} C_{1} z^{4} \\
	+ z^{2} \left(60 C_{0} C_{2} - \frac{320}{3} C_{1}^{2}\right) - 4320 C_{0}^{3}.
\end{multlined}
\end{gather}
\end{subequations}
In this case $g_0'(z)$, $g_1'(z)$ and $g_2'(z)$ have rational primitives and $g_3'(z)$ does not.
The stream functions $\psi_0(z,\bar{z})$, $\psi_1(z,\bar{z})$ and $\psi_2(z,\bar{z})$, given by \eqref{eq:sf-gn}, are hybrid equilibria while $\psi_3(z,\bar{z})$ is not.
This is a 3-link Liouville chain.
The hybrid streamline patterns for the three links are shown in figure~\ref{fig:terminating}.


For any $N\geq 1$ in \eqref{eq:Gamma-N}, the iteration terminates after $N$ steps.
At each step the rational function $g_n'(z)$ is produced by the iterated transformation \eqref{eq:trans-gn} for $n=1,\ldots,N$.
It contains a point vortex at the origin whose circulation is a negative half-integer while the remainder of its circulations satisfy \eqref{eq:circ}. 
Each of these rational functions has a rational primitive.
The circulation of the point vortex at the origin continuously increases by $1$ at each step and the iteration terminates when it reaches the forbidden value $-1/2$.

Each rational function $g_n'(z)$, for $n=1,\ldots,N$, has a corresponding hybrid stream function given by \eqref{eq:sf-gn}. 
The point vortex at the origin in $g_n'(z)$ remains embedded in the hybrid solution as a point vortex with \emph{positive circulation} $N-n-1/2$.
The iteration terminates when this circulation becomes $+1/2$.
The polynomials in the rational functions $g_n'(z)$ above are discussed in another setting by \cite{Duistermaat1986}.

\section{Infinite Liouville chains}\label{sec:infinite}
We construct below three infinite chains of equilibria whose stream functions are given by \eqref{eq:sf-gn}.
The rational functions $g_n'(z)$ in the first Liouville chain we construct are related to the Adler--Moser polynomials \citep{Adler1978}, whereas the rational functions for the second and third Liouville chain are related to the polynomials described in \cite{Loutsenko2004}.
In fact, the example chain discussed in \S\ref{sec:example} is given in terms of the first hierarchy of polynomials due to \cite{Loutsenko2004}; see \S\ref{ss:loutsenko} below.
\subsection{Liouville chain in terms of Adler--Moser polynomials}\label{ss:adlermoser}

\begin{figure}
	\centering\includegraphics[scale=0.5]{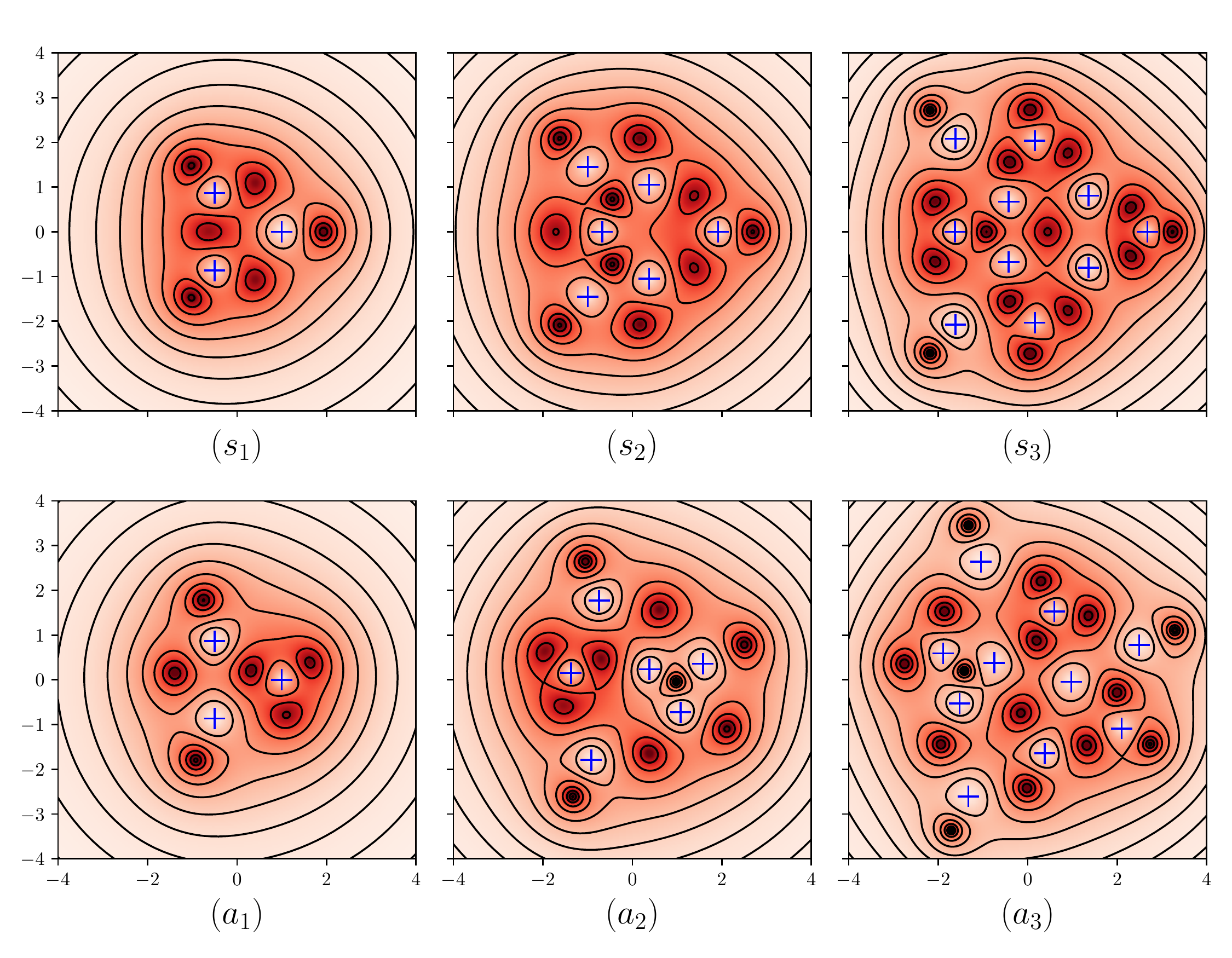}
	\caption{Streamline patterns for a Liouville chain formed from the hierarchy of Adler--Moser polynomials.
		Panels $(s_n)$ and $(a_n)$ are the hybrid stream functions $\psi_n(z,\bar{z})$ given by \eqref{eq:sf-gn} for $n=1,2,3$; the corresponding rational functions $g_n'(z)$ are given by \eqref{eq:AM-ratfun}.
		The panels $(s_n)$ are symmetric configurations whereas $(a_n)$ are asymmetric configurations---they differ in the choice of constants $C_0,\ldots,C_3$ in \eqref{eq:AM-ratfun}.
		The values of these constants are given in table \ref{tab:vals}.
		The point vortex limits of ($s_n$) and ($a_n$) are shown in panels (S1)--(S4) and (A1)--(A4) of figure 3 in \citet{KWCC2020}.}
	\label{fig:adlermoser}
\end{figure}

Successive Adler--Moser polynomials are produced by \eqref{eq:trans-gn} with the seed equilibrium function and twist parameters \citep{KWCC2020} 
\begin{gather}\label{eq:AM-seed}
g_0^{\,\prime}(z)= z^{2}
\quad\text{and}\quad
\alpha_n=-1
\quad\text{for}\quad
n\geq 0.
\end{gather}
Thus $\varGamma=1$ in \eqref{eq:general-seed}.
The first few rational functions $g_n'(z)$ are
\begin{subequations}\label{eq:AM-ratfun}
	\begin{align}
	g_1'(z) &= \frac{\left(z^{3} + 3 C_0\right)^{2}}{z^{2}}, \\
	g_2'(z) &= \frac{\left(z^{6} + 15 C_0 z^{3} + 5 C_1 z- 45 C_0^{2}\right)^{2}}{\left(z^{3} + 3 C_0\right)^{2}}, \\
	g_3'(z) &= \frac{q^2(z)}{\left(z^{6} + 15 C_0 z^{3} + 5 C_1 z - 45 C_0^{2}\right)^{2}},
	\end{align}
	where the polynomial $q(z)$ is
	\begin{gather}
	\begin{multlined}[0.75\linewidth][b]
	q(z) = z^{10} + 45 C_0 z^{7} + 35 C_1 z^{5} + 7 C_2 z^{3} - 525 C_0 C_1 z^{2}\\ 
	+ 4725 C_0^{3} z  + 21 C_0 C_2 - \frac{175}{3}C_1^{2}.
	\end{multlined}
	\end{gather}
\end{subequations}
The Adler--Moser polynomials $p_n(z)$ are then read off from \eqref{eq:AM-ratfun} via the formula 
\begin{gather}
g_n'(z)=\frac{p_{n+1}^2(z)}{p_n^2(z)}\quad\text{for } n\geq 0.
\end{gather}
To compare $p_n(z)$ with the polynomials given in \citet{Adler1978}, redefine their parameters $\tau_2,\tau_3,\ldots$ as $\tau_2=3C_0$, $\tau_3=5C_1$, $\tau_4=7C_2$ and so on. 

The hybrid stream functions follow from substituting \eqref{eq:AM-ratfun} into \eqref{eq:sf-gn} and the corresponding streamline patterns are shown in figure~\ref{fig:adlermoser}.
The limiting cases $A_0=0,\infty$ are stationary pure point vortex equilibria given by $g_0'(z)$ and $g_1'(z)$; the limiting cases $A_1=0,\infty$ are stationary pure point vortex equilibria given by $g_1'(z)$ and $g_2'(z)$, and so on; see the schematic in figure~\ref{fig:schematic}.
The point vortex locations in these limiting patterns are, of course, given by the roots of successive Adler--Moser polynomials and are shown in figure 3 of \citet{KWCC2020}.
The value of the twist parameter $\alpha_n$ in \eqref{eq:AM-seed} is chosen so that at each step the iteration can be continued.

\subsection{Liouville chains in terms of the Loutsenko polynomials}\label{ss:loutsenko}
There are two hierarchies of polynomials described in \citet{Loutsenko2004}, both of which are produced by the iterated transformation \eqref{eq:trans-gn}~\citep{KWCC2020}.
%
The first hierarchy results from the choice of seed equilibrium function and twist parameters
\begin{gather}\label{eq:loutsenko1}
g_0^{\,\prime}(z) = z,
\quad\text{and}\quad
\alpha_n = 
\begin{cases}
-2&\text{ for $n$ even}, \\
-1/2&\text{ for $n$ odd}.
\end{cases}
\end{gather}
This corresponds to setting $\varGamma=1/2$ in \eqref{eq:general-seed}.
We see that $g_0'(z)$ in \eqref{eq:loutsenko1} is actually the input equilibrium function \eqref{eq:g'-link1} with which we started the construction in \S\ref{sec:example}; the twist parameters used there are also the same as in \eqref{eq:loutsenko1}.
In fact, the Liouville chain constructed in \S\ref{sec:example} is given in terms of the first hierarchy of Loutsenko polynomials. 
%
The first few rational functions in this hierarchy, produced by \eqref{eq:trans-gn}, are
\begin{subequations}\label{eq:loutsenko1-ratfun}
	\begin{align}
	g_1'(z) &= \frac{\left(z^{2} + 2 C_0\right)^{4}}{z^{2}}, \\
	g_2'(z) &= \frac{z^{8} + \frac{56}{5}C_0 z^{6} + 56 C_0^{2} z^{4} + 224 C_0^{3} z^{2} + 7 C_1 z - 112 C_0^{4}}{\left(z^{2} + 2 C_0\right)^{2}}, \\
	g_3'(z) &= \frac{\left(z^{7} + 14 C_0 z^{5} + 140 C_0^{2} z^{3} + 5 C_2 z^{2} - 280 C_0^{3} z + 10 C_0 C_2 - \frac{35}{2} C_1\right)^{4}}
	{\left(z^{8} + \frac{56}{5} C_0 z^{6} + 56 C_0^{2} z^{4} + 224 C_0^{3} z^{2} + 7 C_1 z - 112 C_0^{4}\right)^{2}}.
	\end{align}
\end{subequations}
The hybrid equilibria follow from substitution of \eqref{eq:loutsenko1-ratfun} into \eqref{eq:sf-gn}.
Streamline patterns for the Liouville links $n=0,1,2$ in this chain are shown in figure \ref{fig:loutsenko1}.

Polynomials $p_n(z)$ are read off from the rational function $g_n'(z)$ using the formula 
\begin{gather}\label{eq:loutsenko1-identify}
g_n^{\,\prime}(z) =  
\begin{cases}
\dfrac{p_{n+1}(z)}{p_n^2(z)}&\text{ for $n$ even}, \\[3ex]
\dfrac{p_{n+1}^4(z)}{p_n^2(z)}&\text{ for $n$ odd}.
\end{cases}
\end{gather}
To compare $p_n(z)$ with the polynomials described in \citet{Loutsenko2004}, the branch $i \le 0$ in his notation, first rename the polynomials $p_n(z)$ according to $p_n\to p_{-(n+1)/2}$ for $n$ odd and $p_n\to q_{-n/2}$ for $n\geq 2$ and even. 
We can make the identification after redefining his parameters $\tau_{-1},t_{-2},\tau_{-2},\ldots$ as $\tau_{-1}=2C_0$, $t_{-2}=7C_1$, $\tau_{-2}=5C_2$ and so on.


\begin{figure}
	\centering\includegraphics[scale=0.5]{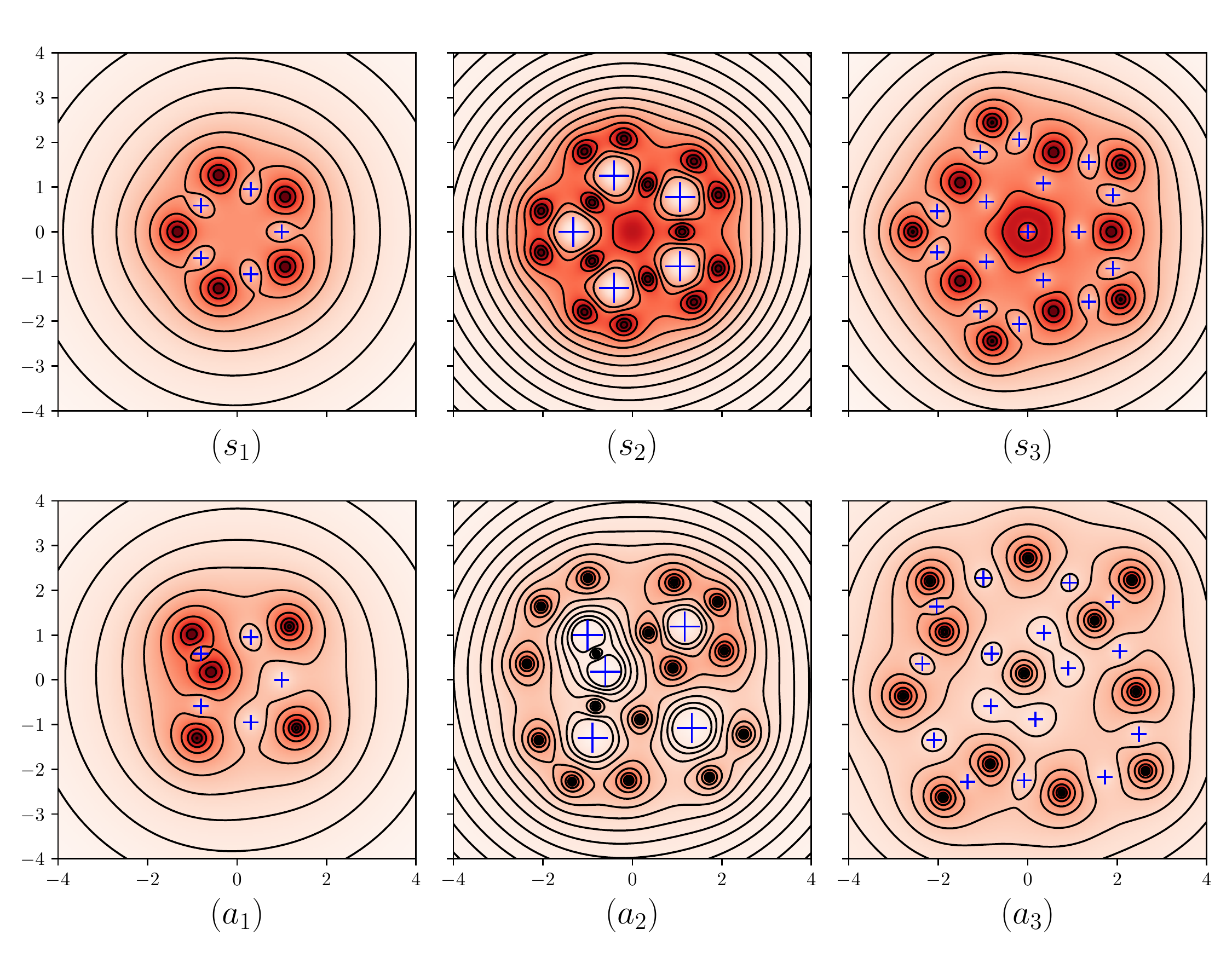}
	\caption{Symmetric $(s_n)$ and asymmetric $(a_n)$ streamline patterns of the hybrid stream functions \eqref{eq:sf-gn} for $n=1,2,3$ given in terms of the second Loutsenko hierarchy \eqref{eq:loutsenko2-ratfun}. 
	The values of the constants used here are given in table \ref{tab:vals}.
	The corresponding point vortex limits are shown in figure 5 of \citet{KWCC2020}.}
	\label{fig:loutsenko2}
\end{figure}
The second hierarchy due to Loutsenko is produced using \eqref{eq:trans-gn} with the seed equilibrium function and twist parameters taken to be
\begin{gather}\label{eq:loutsenko2}
g_0^{\,\prime}(z) = z^4,
\quad\text{and}\quad
\alpha_n = 
\begin{cases}
-1/2&\text{ for $n$ even}, \\
-2&\text{ for $n$ odd}.
\end{cases}
\end{gather}
The first few rational functions are 
\begin{subequations}\label{eq:loutsenko2-ratfun}
	\begin{align}
	g_1'(z) &=  \frac{z^{5} + 5 C_0}{z^{2}}, \\
	g_2'(z) &=  \frac{\left(z^{5} + 4 C_1 z - 20 C_0\right)^{4}}{\left(z^{5} + 5 C_0\right)^{2}}, \\
	g_3'(z) &=  \frac{q(z)}{\left(z^{5} + 4 C_1 z - 20 C_0\right)^{2}},
	\end{align}
	where the numerator polynomial $q(z)$ in $g_3'(z)$ is
	\begin{multline}
	q(z) = z^{16} + \frac{176}{7}C_1 z^{12} - 160 C_0 z^{11} + 352 C_1^{2} z^{8} - \frac{42240}{7} C_0 C_1 z^{7}  \\
	+ 35200 C_0^{2} z^{6} + 11 C_2 z^{5} - 2816 C_1^{3} z^{4} + 28160 C_0 C_1^{2} z^{3}  \\
	- 140800 C_0^{2} C_1 z^{2} + 352000 C_0^{3} z - \frac{2816}{5} C_1^{4} + 55 C_0 C_2.
	\end{multline}
\end{subequations}
We can read off polynomials $p_n(z)$ from \eqref{eq:loutsenko2-ratfun} using the formula
\begin{gather}\label{eq:loutsenko2-identify}
g_n^{\,\prime}(z) =  
\begin{cases}
 \dfrac{p_{n+1}^4(z)}{p_n^2(z)} &\text{ for $n$ even}, \\[3ex]
 \dfrac{p_{n+1}(z)}{p_n^2(z)} &\text{ for $n$ odd}.
\end{cases}
\end{gather}
To compare $p_n(z)$ with the polynomials from the branch $i \geq 0$ in \citet{Loutsenko2004}, rename $p_n(z)$ according to $p_n\to p_{n/2}$ for $n\geq 0$ even and $p_n\to q_{(n+1)/2}$ for $n$ odd; and redefine his parameters as $t_{1}=5C_0$, $\tau_2=4C_1$ and so on.

Figure~\ref{fig:loutsenko2} shows the streamline patterns for the hybrid equilibria obtained from \eqref{eq:loutsenko2-ratfun}, for the cases $n=1,2,3$. 
The corresponding limiting point vortex patterns of these Liouville links are shown in figure~5 of \citet{KWCC2020}.
It is seen in figure~\ref{fig:loutsenko2} that the inter-streamline distance alternately increases and decreases as we move up the hierarchy. 
The total circulation of the hybrid equilibria is calculated in appendix \ref{app:circ}.
From \eqref{eq:Gamma-hyb}, we see that for any finite $A_n$ the total circulation of the hybrid solution is simply given in terms of the net circulation of the underlying point vortex equilibrium, $\varGamma_\text{hyb}=-(\varGamma_\text{pv}+1)$ since $\varGamma_\text{pv}>0$.
It is easily calculated from \eqref{eq:loutsenko1} and \eqref{eq:loutsenko2} that $\varGamma_\text{pv}$ oscillates up and down as we move up the hierarchy, which is the reason for the alternating streamline patterns.
Although this is true also for figure~\ref{fig:loutsenko1}, the pattern is not as easily visible there.

\section{Summary and future directions}\label{sec:summary}
A large class of exact solutions of the 2D Euler equation in the form of hybrid vortical equilibria has been derived. 
These solutions, named Liouville chains, are given by simple analytical expressions involving elementary functions.
The individual links in the chain, called Liouville links, are hybrid vortex equilibria parametrized by a positive real parameter $A$ and comprise point vortices embedded in an ambient Liouville-type field of smooth vorticity.
Every Liouville link connects two pure point vortex equilibria that emerge at each end of the parameter range: $A\to 0$ and $A\to\infty$. 
We might think of the hybrid equilibria as ``extrapolating'' between the two pure point vortex equilibria. 
Liouville chains can be finite or infinite, with a possible twist needed to successfully produce a next link in the chain. 
Among other examples, three infinite Liouville chains have been presented explicitly: one associated with the Adler--Moser polynomials, and two others associated with polynomial hierarchies found by \citet{Loutsenko2004}.

The transformation between stationary point vortex equilibria presented in \cite{KWCC2020} links the two limiting cases of the hybrid equilibria.
The hybrid equilibria are rotational solutions that connect two distinct irrotational solutions of the steady Euler equation, continuously deforming one into the other as the parameter $A$ is varied.
The present study shows that it can be non-trivial to choose an appropriate function $h'(z)$ in \eqref{eq:liouville-sol} which leads to a meaningful steady solution of the Euler equation. 
Future investigations, with various other choices of $h'(z)$, could reveal a variety of simple and explicit solutions which are of the hybrid-type presented here.

Although the hybrid equilibria contain free parameters in the form of the integration constants $C$, the physical meaning---if any---of these parameters has not been investigated.
However, the presence of these free parameters can be exploited to model different types of flows. 
\cite{Meiburg1991} and \cite{Newton1991a} studied the effect of viscosity on shear flows by taking the parameter in the original Stuart solution \citep{Stuart1967} to depend on viscosity.
\cite{Fraenkel2008} considered the same question in a more general setting by taking the Stuart solutions as the initial condition for the unsteady Navier--Stokes equation.
One might also consider the current solutions as the initial condition for the unsteady Euler equation and, by allowing the parameters  to depend on time,  investigate the resulting dynamics.

\citet{PW1981} considered the question of stability of the original Stuart vortex row \citep{Stuart1967}.
The nonlinear stability of Stuart vortices was examined by \citet{Holm1986}, who found a range of parameter values for which stable solutions exist.
See \citet{Friedlander1999} for a general introduction to stability of solutions to the Euler equation.
All these stability results will likely be modified if point vortices are present in the flow, and the stability of hybrid solutions with multiple point vortices is an important question that needs to be addressed.
Some recent progress for a single point vortex in a perturbed background vorticity field is reported in \citet{Alex2019}.

The effects of compressibility on the incompressible Stuart vortex solution was considered by \citet{Meiron2000}, who used Rayleigh-Jansen expansions for analytical studies of low-Mach number flow.
The classical K\'arm\'an point vortex street has been extended to weakly compressible flow recently by \citet{CK2017a}, who also discuss and clarify the general force-free condition required for a point vortex to be in equilibrium in weakly compressible flows.
The weakly compressible counterpart of the hollow vortex street \citep{CroGreen} is studied by \citet{CK2017b}. 
This recent activity on weakly compressible flows with embedded vortices gives analytical and numerical evidence for the existence of smooth transonic flows.
The weakly compressible counterparts of the hybrid equilibria studied here (unlike the above cases, these are not periodic) are a natural object of study in this respect.

It is possible to make different choices for the vorticity function $V(\psi)$ in \eqref{eq:steady-vorteq} and obtain corresponding steady solutions to the Euler equation.
The choice $V(\psi)=\sinh\psi$ gives the sinh-Poisson equation considered by \cite{Mallier1993}, whose periodic solution consists of an alternating row of counter-rotating vortices.
Using more abstract methods than the ones presented here, it has been shown by \cite{Bartsch2010} that the sinh-Poisson equation in a bounded domain  has a limiting case where the vorticity concentrates into delta-distributions.
In fact, their paper utilises the Liouville equation whose explicit solutions are considered here.
The class of all functions $V$ with smooth vorticity, which concentrate into point vortices in some appropriate limits, does not appear to have been investigated in detail.

There is a large mathematical literature on singular solutions of the Liouville equation in bounded domains. 
See for instance \citet{delPino2010,Ma2001}.
Moreover, \cite{Gustafsson1979} and \cite{Richardson1980} have independently shown that the Hamiltonian for a single point vortex in a simply connected domain also obeys the Liouville equation with the boundary condition that the Hamiltonian is infinite on the boundary.
\cite{Crowdy2006} has shown that the same idea extends to point vortex motion in a simply connected domain on the surface of a sphere.

The proofs presented in this paper use local analyses, but an alternate treatment borrowing ideas from mathematical physics can be given. 
Some of the connections to mathematical physics may be seen from the appearance of the Adler--Moser polynomials \citep{Adler1978} and the Loutsenko polynomials \citep{Loutsenko2004}.
These polynomials are not new in the context of purely point vortex equilibria; for a detailed discussion see for example \cite{Aref2003,Aref2007,Clarkson2009,KWCC2020}.
But the appearance of these polynomials in the context of rotational flows and specifically the Liouville-type equation \eqref{eq:liouville-type} calls for further investigation.

Several other avenues for extending the solutions derived here present themselves. 
The original solutions due to \cite{Stuart1967} are singly-periodic solutions and we can ask if there are singly-periodic hybrid equilibria.
Stuart vortex solutions have been extended to a sphere \citep{Crowdy2004b} and a torus \citep{Sakajo2019}.
The existence of more general hybrid equilibria on these compact surfaces, consisting of greater numbers of point vortices, is an open question.
Extensions of Stuart vortex solutions to the rotating sphere have also been considered recently in the context of geophysical applications, to study large scale planetary structures such as gyres \citep{CK2019} and polar vortices \citep{CCKW2020}.
The utility of the much richer set of hybrid solutions presented here, in such applications, needs further examination.
We have already made progress in some of these directions and our results will be reported elsewhere.

{\par\noindent\bfseries Declaration of Interests.} The authors report no conflict of interest.
                                                                                        
\appendix

\section{Net circulation of the flow}\label{app:circ}
In this appendix, we calculate the total circulation of the hybrid equilibria,
\begin{equation}\label{eq:circ-def}
	\varGamma_\text{hyb} 
	= \lim_{r \to \infty} \oint\limits_{|z|=r} (u\, \mathrm dx + v\, \mathrm dy)
	= \lim_{r \to \infty} \Real \oint\limits_{|z|=r} (u-\mathrm iv)\, \mathrm dz,
\end{equation}
in terms of the net circulation 
$\varGamma_\text{pv} = \sum_{j=1}^M \varGamma_j$ of the input point vortex equilibrium represented by \eqref{eq:g'}. 

Let us first consider $\Gamma_\text{hyb}$ in the point vortex limits $A \to 0,\infty$, which have the complex potentials $G,F$ given in \eqref{eq:limit1}. 
As $|z| \to \infty$, the rational function ${g'(z) \sim z^{2\varGamma_\text{pv}}}$. 
From \S\ref{ss:proof1} we know that the primitive $g(z)$ is rational so that ${\varGamma_\text{pv} \ne -1/2}$ and hence ${g(z) \sim z^{2\varGamma_\text{pv}+1}}$. This yields the asymptotics
\begin{gather}
	G(z) \sim \frac{\varGamma_\text{pv}}{2\upi\mathrm{i}}\log z
	\qquad\text{and}\qquad
	F(z) \sim -\frac{(\varGamma_\text{pv}+1)}{2\upi\mathrm{i}}\log z,
\end{gather}
implying that the total circulations are 
\begin{gather}\label{eq:circ-limits}
	\varGamma_\text{hyb} = 
	\begin{cases}
	\varGamma_\text{pv}\quad &\text{at }A= 0,\\[1ex]
	-(\varGamma_\text{pv}+1)\quad &\text{at }A=\infty.
	\end{cases}
\end{gather}
Note that when comparing $\varGamma_\text{hyb}$ at $A=\infty$ with the circulations of the point vortices in the equilibrium function $\widehat{g}'(z)$ introduced via the transformation \eqref{eq:trans-g}, we need to further scale by the parameter $\alpha$.

Turning now to true hybrid equilibria with $0 < A < \infty$, there is no complex potential and instead we must use the formula \eqref{eq:hybrid-sf} for the hybrid stream function. 
As $|z|\to\infty$ we have, as earlier, $g'(z)\sim z^{2\varGamma_\text{pv}}$ and $g(z)\sim z^{2\varGamma_\text{pv}+1}$.
We find after some calculation that the behavior of the hybrid stream function now depends on the sign of $\varGamma_\text{pv}$, 
\begin{gather}
	\psi\sim
	\begin{cases}
	-\dfrac{\varGamma_\text{pv}}{2\upi}\log|z|+ O\left(\dfrac{1}{|z|}\right)\quad &\text{if }\varGamma_\text{pv}<0, \\[2ex]
	\dfrac{(\varGamma_\text{pv}+1)}{2\upi}\log|z| + O\left(\dfrac{1}{|z|}\right)\quad &\text{if }\varGamma_\text{pv}>0.		
	\end{cases}
\end{gather}
The total circulation can be read off as 
\begin{gather}\label{eq:Gamma-hyb}
	\varGamma_\text{hyb} = 
	\begin{cases}
	\varGamma_\text{pv}\quad &\text{if }\varGamma_\text{pv}<0, \\[1ex]
	-(\varGamma_\text{pv}+1)\quad &\text{if }\varGamma_\text{pv}>0.
	\end{cases}
\end{gather}
In particular, $\varGamma_\text{hyb}$ is independent of $A$ for $0 < A < \infty$, but is discontinuous at $A=0$ (if $\varGamma_\text{pv} > 0$) or $A=\infty$ (if $\varGamma_\text{pv} < 0$). 
This discontinuity can be interpreted as some of the negative Liouville-type vorticity ``leaking out to infinity''. 
Mathematically, it occurs because the limit $r \to \infty$ in \eqref{eq:circ-def} does not commute with the relevant limit $A \to 0$ or $A \to \infty$. 
Note that if the value of $b$ differs from $-8\upi$, then the circulations $\varGamma_\text{pv}$ and $(\varGamma_\text{pv}+1)$ in the above formulas are scaled by $-8\upi/b$.

\section{Conventions and values for the parameters}\label{app:convention}
\begin{table}
	\centering
	\begin{tabular}{rr|l|l|l|l|l} \\ \toprule
		\multicolumn{2}{c|}{Figure} & $A$ &$C_0$ & $C_1$ & $C_2$ & $C_3$ \\ \midrule
		\ref{fig:terminating} & $(s_{3,1})$ & $A_1=3$ & $1$ &     & & \\ 
		& $(s_{3,2})$ & $A_2=2 \times 10^{-1}$ & $1$ & $0$ &     & \\
		& $(s_{3,3})$ & $A_3=1.5 \times 10^{-2}$ & $1$ & $0$ & $0$    & \\
		& $(a_{3,1})$ & $A_1=3$ & $1+\mathrm{i}$ &                & &\\
		& $(a_{3,2})$ & $A_2=2 \times 10^{-1}$ & $1+\mathrm{i}$ & $10\mathrm{i}$ &      & \\
		& $(a_{3,3})$ & $A_3=1.5 \times 10^{-2}$ & $1+\mathrm{i}$ & $10\mathrm{i}$ & $-8$ & \\ \midrule
		\ref{fig:adlermoser} & $(s_1)$ & $A_1=1$ & $-1/3$ & $-1$ &  &  \\ 
		& $(s_2)$ & $A_2=5 \times 10^{-2}$ & $-1/3$ & $-1$ & $20$ &  \\
		& $(s_3)$ & $A_3=3 \times 10^{-3}$ & $-1/3$ & $-1$ & $20$ & $80$ \\
		& $(a_1)$ & $A_1=1$ & $-1/3$ & $2-\mathrm{i}$ & & \\
		& $(a_2)$ & $A_2=5 \times 10^{-2}$ & $-1/3$ & $2-\mathrm{i}$ & $8-8\mathrm{i}$ &  \\
		& $(a_3)$ & $A_3=3 \times 10^{-3}$ & $-1/3$ & $2-\mathrm{i}$ & $8-8\mathrm{i}$ & $40 + 120\mathrm{i}$ \\ \midrule
		\ref{fig:loutsenko2} & $(s_1)$ & $A_1=3$ & $-1/5$ & $0$ & & \\ 
		& $(s_2)$ & $A_2=10^{-2}$ & $-1/5$ & $0$ & $0$ & \\
		& $(s_3)$ & $A_3=5 \times 10^{-2}$ & $-1/5$ & $0$ & $0$ & $0$ \\
		& $(a_1)$ & $A_1=3$ & $-1/5$ & $(3+\mathrm{i})/2$ & &  \\
		& $(a_2)$ & $A_2=10^{-2}$ & $-1/5$ & $(3+\mathrm{i})/2$ & $1000-2000\mathrm{i}$ &  \\
		& $(a_3)$ & $A_3=5 \times 10^{-2}$ & $-1/5$ & $(3+\mathrm{i})/2$ & $1000-2000\mathrm{i}$ & $100+100\mathrm{i}$ \\ \midrule
		\bottomrule
	\end{tabular}
	\caption{
		Values of constants used in figures \ref{fig:terminating}, \ref{fig:adlermoser} and \ref{fig:loutsenko2}.
		The constants $C_n$ are calculated according to the convention discussed in appendix \ref{app:convention}.}
	\label{tab:vals}
\end{table}
In our discussions we have considered a particular integral $h(z)$ of $h'(z)$, for example in the proof sections \S\ref{ss:proof1} and \S\ref{ss:proof2}.
Looking at a particular integral $g(z)$ of the function $g'(z)$, the general integral involves an integration constant $C$.
Taking this into consideration, we have written a general integral of $g'(z)$ as $g(z)+C$.
These choices are made according to a set convention that we follow.
The convention for $C$ that we now describe is closely related to that adopted in \citet{KWCC2020}.

If $N(z)/D(z)$ is a rational primitive of $g'(z)$ for some polynomials $N(z)$ and $D(z)$, then by polynomial long division
\begin{align}\label{eq:N/D}
\frac{N(z)}{D(z)} = P(z) + \frac{R(z)}{D(z)},
\end{align}
where here $P(z)$ and $R(z)$ are polynomials and the degree of $R(z)$ is strictly less than that of $D(z)$.
We define 
\begin{align}\label{eq:cn}
g(z) = P(z) - P(0) + \frac{R(z)}{D(z)}.
\end{align}
This is equivalent to setting the constant term in \eqref{eq:N/D} equal to zero.
Adding in the constant $C$ explicitly, the general antiderivative of $g'(z)$ can be written as $g(z)+C$.
The functions $h'(z)$ and $h(z)$ are then given by \eqref{eq:h'-g'}.

We now turn to a description of the convention for the constant $\widehat{E}$.
The definition \eqref{eq:g'-repeat} of the rational function $g'(z)$ shows its numerator and denominator polynomials to be monic.
On the other hand, the rational function $g(z)+C$ will not have monic numerator and denominator polynomials in general.
We treat $\widehat{E}$ in \eqref{eq:trans-g} as a constant and choose its value so that $\widehat{g}'(z)$ also has monic numerator and denominator polynomials. 
The convention for the constants $E_n$ appearing in \eqref{eq:trans-gn} is also similar.
Accordingly, we choose the constant $E_{n}$ such that $g_{n}'(z)$ consists of monic numerator and denominator polynomials for all $n\geq 0$.

A quick note here about comparing the notation of the present paper with that in \citet{KWCC2020}. 
Since that paper deals exclusively with pure point vortex equilibria, adding constants to the complex potential leaves the velocity field undisturbed and the parameter $A$ is a constant with no physical significance there.
While comparing the two papers it is helpful to respectively translate the $A,\widehat{A},A_n$ and $h(z),h'(z)$ that appear in \citet{KWCC2020} to the $E,\widehat{E},E_n$ and $g(z),g'(z)$ that appear in the current paper.

The rational functions presented in this paper were all obtained using computer algebra packages. 
The computer algebra was performed using the SymPy Python library and cross-checked using Mathematica.
Different computer algebra packages can produce different rational functions as primitives of a given rational function. 
The rational functions produced at the end can differ significantly, especially when the procedure is iterated like in Liouville chains.
It is therefore important to follow a set convention for the integration constants and other parameters, as described here.

The constants $a$ and $b$ in the hybrid stream function \eqref{eq:liouville-sol} should satisfy the constraint $ab<0$.
We choose $b=-8\upi$, so that the circulation of the embedded positive point vortices in the hybrid solution is scaled by $1$; see the end of \S\ref{ss:proof2}.
We then set $a=1/4\upi$ so that $ab=-2$ and the argument in the logarithm of \eqref{eq:liouville-sol} simplifies.
From \eqref{eq:vort-sf} and \eqref{eq:liouville-type}, the sign of the background vorticity is determined by the sign of $-a$; hence our choice ensures negative Liouville-type vorticity.
This scaling is consistent with the scaling used in \citet{KWCC2019}.
The values of the parameters $A_n$ and $C_n$ used to produce the figures are given in table~\ref{tab:vals}, unless specified in the figure itself.
Successive streamline contours are separated by a constant value of the stream function in all of the figures, but this value can vary across figures.

\bibliographystyle{jfm}
\bibliography{citations-JFM}

\end{document}